\documentclass[aps,pre,twocolumn,reprint,longbibliography]{revtex4-2}
\usepackage{graphicx,color}
\usepackage[colorlinks=true,linkcolor=blue,citecolor=blue,urlcolor=blue]{hyperref}
\usepackage{amsmath}
\usepackage{amsthm}
\usepackage{amssymb}
\usepackage{physics}
\usepackage{amsfonts}
\usepackage{tabularx}
\usepackage{comment}
\usepackage[normalem]{ulem}
\usepackage{soul}

\newcommand{\be}{\begin{equation}} 
\newcommand{\ee}{\end{equation}} 
\newcommand{\bea}{\begin{eqnarray}} 
\newcommand{\eea}{\end{eqnarray}}

\newcommand{\imp}{\qq{$\Rightarrow$}}
\newcommand{\Z}{\mathcal{Z}}

\begin{document}
\title{Position-Momenta Uncertainties  in   Classical Systems}
\author{Dipesh K. Singh}
\email{dks20ms176@iiserkol.ac.in}
\author{P. K. Mohanty}
\email{pkmohanty@iiserkol.ac.in}
\affiliation {Department of Physical Sciences, Indian Institute of Science Education and Research Kolkata, Mohanpur, 741246, India.}
\date{\today}

\begin{abstract}
We design a thermal bath that preserves the conservation of a system’s angular momentum or allows it to fluctuate around a specified nonzero mean while maintaining a Boltzmann distribution of energy in the steady state. We demonstrate that classical particles immersed in such baths exhibit position-momentum uncertainties with a strictly positive lower bound proportional to the absolute value of the mean angular momentum. The proportionality constant, $c$, is dimensionless and does not  depend explicitly on  the system’s parameters. Remarkably, while $c$ is universally bounded by unity, it attains the exact value $c=\frac12$ for particles in central potentials. 
\end{abstract}
\maketitle

\section{\label{intro:level1}Introduction}
The Heisenberg uncertainty principle \cite{Heisenberg1927} marks a fundamental departure of quantum mechanics from classical physics, imposing intrinsic limits on the simultaneous measurement of conjugate variables such as position and momentum. Given that both quantum mechanics and classical statistical physics share a probabilistic framework, natural questions arise—whether macroscopic classical realism can emerge from quantum principles  \cite{Preskill2013} and whether classical systems subject to thermal fluctuations can exhibit uncertainty relations analogous to quantum bounds.
Recent studies have shown that coarse-grained measurements of quantum systems can give rise to classical Newtonian laws directly from the Schrödinger equation \cite{Kofler2007}. Similarly, quantum particles in arbitrary potentials with passive dissipation can lead to classical Langevin equations \cite{Ford1988}. Concurrently, advances in statistical mechanics and stochastic thermodynamics have revealed that classical systems experiencing thermal fluctuations can exhibit uncertainty relations reminiscent of quantum constraints. For instance, stochastic thermodynamics reveals fluctuation-dissipation relations \cite{Marconi2008} and thermodynamic uncertainty principles \cite{Barato2015, Gingrich2016,Shiraishi2016} that mirror quantum bounds. In nonequilibrium systems, interplay between noise and dissipation leads to thermodynamic uncertainty relations that constrain the precision of nonequilibrium fluctuations \cite{Maes2003, Jarzynski1997, Horowitz2020}. 
 
 The connection between quantum fluctuations and Brownian motion was first recognized in 1933 by R. F\"urth 
 \cite{Peliti2023}. Recently, it has been proposed that quantum-like entanglement can emerge between interacting Brownian particles \cite{Allahverdyan2000}, a phenomenon later verified experimentally \cite{Ciliberto2024}. In this experiment, two coupled electric circuits mimicking Brownian particles exhibited entanglement-like behavior, attributed to the fact that their positions and coarse-grained velocities obey classical uncertainty relations. These findings raise a fundamental question: Can classical systems exhibit genuine uncertainty bounds akin to those in quantum mechanics?

In this work, we show that classical particles in a thermal bath—where angular momentum is either strictly conserved or fluctuates around a non-zero mean—obey a lower bound on position-momentum uncertainties proportional to the mean angular momentum. The proportionality constant $c$ is dimensionless  
and always $c\le1$, mirroring the numerical prefactors often seen in quantum uncertainty relations; remarkably, for central potentials, it takes the exact value $\frac12$. This result extends the connection between classical statistical mechanics and quantum-like uncertainty, suggesting that certain features of quantum indeterminacy may have classical counterparts under appropriate constraints \cite{Barato2015, Horowitz2020}.

\begin{figure}[t]
    \centering
    \includegraphics[width=\linewidth]{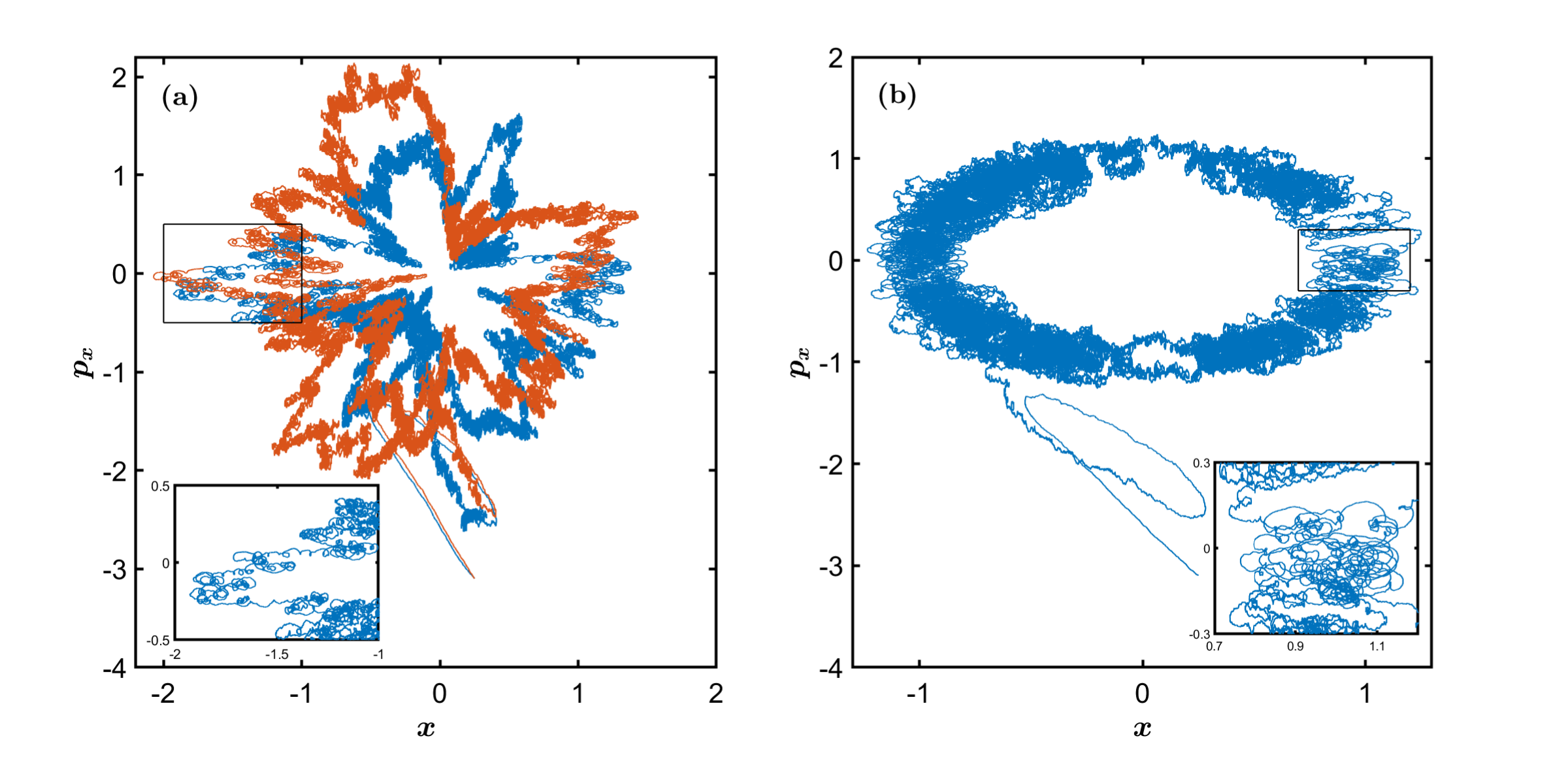}\vspace*{-.3cm}
    
    \caption{Sample paths for  $V(r)=r^\alpha/\alpha$ at $T=10^{-3}$
     with  $\ev{p_\phi}=1$ and $\gamma=2.$ (a) $\alpha=2$ (two sample paths) (b) $\alpha=2.1.$ The insets show the spiraling nature of trajectories.} \vspace*{-.3cm}
     
    \label{fig:noncons_samplepaths}
\end{figure}

We evaluate three distinct scenarios in two dimensions (2D). First,   classical particles in   central potentials   which keeps their   angular momentum  $p_\phi=\ell$ conserved.   We show  that irrespective of the form of 
the potential, they  obey  an uncertainty relation, 
\be
\Delta x \Delta p_x   = \Delta y \Delta p_y \ge \frac{|\ell|}{2},
\label{eq:ur1}
\ee
when placed  in a thermal bath  that  respects  conservation of angular momentum, $p_\phi=\ell.$ Here,  uncertainty   in measurement of  an observable  $A$  is   the   standard  deviation  $\Delta A =  \sqrt{ \ev{A^2} - \ev{A}^2 }$ in the steady state. 
Next, in the true spirit of the generalized Gibbs ensemble \cite{Gibbs2010},  we devise a Langevin bath  that allows    $p_\phi$  to fluctuate around a desired mean $\ev{p_\phi}$  while  ensuring  that  the system reaches  an equilibrium state having a Boltzmann distribution at temperature  $T$, and show that the  observables  obey  a similar  uncertainty relation,  with angular momentum $\ell$ replaced by  its average  
\be
\Delta x \Delta p_x   = \Delta y \Delta p_y \ge \frac{\abs{\ev {p_\phi}}}{2}.
\label{eq:ur2}
\ee
In both cases, the individual uncertainties  of $x,y$ and  $p_{x,y}$ depend on  the parameters of the system, i.e.,  parameters   of the  potential $V(r),$  and  they  are increasing functions of temperature. However,  the lower  bound  of  the  uncertainty-product of conjugate variables, obtained in the  $T\to0$ limit, depends only  on 
the angular  momentum $p_\phi=\ell$ (when conserved)  or  its mean $\ev {p_\phi}$  (when  $p_\phi$ fluctuates). 

In a generic non-radial potential $V(r,\phi),$  the angular  momentum  of  the  particle  is not conserved.  We use the same thermal bath considered before, which allows $p_\phi$ fluctuations about a desired mean. The uncertainty relation  in this case is  more  general; the lower bound is proportional to $\abs{\ev {p_\phi}}$  with a dimensionless prefactor $c$ which can not exceed $\frac12,$
\be
 \Delta x \Delta p_x\ge c {\abs{\ev {p_\phi}}}; ~~
\Delta y \Delta p_y \ge c {\abs{\ev {p_\phi}}}.
\label{eq:ur3}
\ee
Here, in general,   $\Delta x\Delta p_x\neq\Delta y\Delta p_y$ but their lower bound is  the same.

Our findings align with recent studies on thermodynamic bounds in stochastic systems \cite{Maes2003, Jarzynski1997} and contribute to the broader discussion on the interplay between noise and  symmetry in classical systems in equilibrium. They  further extend the growing body of work on thermodynamic uncertainty relations \cite{Seifert2012, Gingrich2016,Shiraishi2016, Hyeon2017,Macieszczak2018}, which have revealed deep connections between information, fluctuations, and measurement precision in nonequilibrium systems. They complement recent studies on emergent quantum-like behavior in classical stochastic systems \cite{Nelson1966, deOliveira2024}.

Thermodynamic uncertainty relations \cite{Dieball2023, Kwon2022} or statistical bounds \cite{Sabbagh2024}  are not new  to statistical physics, but their presence, particularly in equilibrium,  carries significant implications for our understanding of the quantum-classical divide.  It is true that the position-momenta uncertainty relations are built into the foundations of quantum mechanics, and while both statistical physics and quantum mechanics share a probabilistic framework, it is still possible to realize single particle trajectories, like in Figs. \ref{fig:noncons_samplepaths} and \ref{fig:distribution_sho}, using stochastic dynamics and different types of thermal baths in statistical physics. However, our work suggests that certain phenomena traditionally regarded as uniquely quantum mechanical may  also arise in  classical statistical systems, emerging through the interplay of thermal fluctuations and dynamical constraints.

 The article is structured as follows. We present the conserved and non-conserved angular momentum baths along with the example of power law potentials in Secs. \ref{sec:consL} and \ref{sec:nonconsL} respectively. A rich discussion on the nature of the constructed baths follows in Sec. \ref{sec:discussion}. We work with non-central potentials in Sec. \ref{sec:noncentral} and a brief overview of higher dimensions is presented in Sec. \ref{sec:higherD}. The conclusion with a few remarks is given in Sec. \ref{sec:conc}.

\section{\label{sec:consL} Conserved Angular Momentum}
Consider the Hamiltonian in two dimensions, with coordinates $(r,\phi)$ and momenta $(p_r, p_\phi),$
\begin{equation}
    H=\dfrac{p_r^2}{2}+\dfrac{p_\phi^2}{2r^2}+V(r).
\end{equation}
Since $\phi$ is cyclic, the angular momentum $p_\phi=\ell$  is conserved. We use a radial Langevin bath which respects this conservation law. The equations of motion are now given by
\begin{equation}
    \dot{r}=p_r; \dot{p}_r=-\pdv{V}{r}+\dfrac{\ell^2}{r^3}-\gamma p_r+\sqrt{\gamma T}\xi(t); \dot{\phi}=\dfrac{\ell}{r^2},
    \label{eq:2dLangevin}
\end{equation}
where $\gamma>0$ and $\xi(t)$ is a Gaussian white noise with $\ev{\xi(t)}=0$ and $\ev{\xi(t)\xi(t')}=2\delta(t-t')$. We use units where the mass and Boltzmann constant are unity $(m=k_B=1)$; we define $\beta\equiv T^{-1}$. 

For any $T\ne0,$ the Fokker-Planck equation corresponding to the Langevin dynamics  \eqref{eq:2dLangevin} is
\begin{equation}
    \pdv{P}{t}=[H,P]+\pdv{p_r}(\gamma p_rP+\gamma T\pdv{P}{p_r}),
\end{equation}
with $[~,~]$ being the Poisson bracket. The above equation admits a steady state given by the Boltzmann distribution $P(r,p_r,p_\phi=\ell)=\frac{e^{-\beta H}}{\mathcal{Z}_\ell(\beta)}$, where the partition function is
\begin{equation}
    \mathcal{Z}_\ell(\beta)=2\pi\sqrt{\dfrac{2\pi}{\beta}}\int_{0}^{\infty}\dd{r}\exp(-\beta\qty(\dfrac{\ell^2}{2r^2}+V(r))).\label{conditioncons}
\end{equation}
Of course, $\mathcal{Z}_\ell(\beta)$ must be finite, as required for a valid probability density function (PDF)  (see Appendix \ref{app:FP} for details). 

From the symmetries of the  Hamiltonian, it is clear that $\ev{x^n}=\ev{y^n}, \ev{p_x^n}=\ev{p_y^n}\forall n$ and they vanish  when $n$ is  odd. Further, $\ev{x^2}=\frac12 \ev{r^2}$ and from the relations $\frac{1}{2}(p_x^2+p_y^2)=\frac{1}{2}(p_r^2+\frac{\ell^2}{r^2})$ and $\frac{1}{2}\ev{p_r^2}=\frac{T}{2}$, we get $\ev{p_x^2}=\frac12\qty(T+\ell^2\ev{\frac{1}{r^2}}).$ 
Therefore,
\begin{equation}
    \Delta x\Delta p_x=\Delta y\Delta p_y=\frac12 \sqrt{\ev{r^2}\qty(T+\ell^2\ev{1/r^{2}})}. \label{eq:fullUncertainty}
\end{equation}
 We know that $T\ev{r^2}\geq0$ using which, it is easy to see that  $\sqrt{\ev{r^2}\qty(T+\ell^2\ev{1/r^{2}})}\geq\sqrt{\ell^2\ev{r^2}\ev{1/r^{2}}}$ with the equality holding as $T\to0$. Provided the steady state exists and $\ev{r^2}$ is finite (and hence $\ev{\frac{1}{r^2}}$ is also finite), the Cauchy-Schwarz inequality ensures  $\ev{r^2}\ev{1/r^2}\geq1$ and then Eq. \eqref{eq:fullUncertainty} leads us to  the uncertainty relations, 
\begin{equation}
    \Delta x\Delta p_x=\Delta y\Delta p_y\geq\dfrac{\abs{\ell}}{2}. \label{uncertaintycons}
\end{equation}
From the above relations, it is clear the there are positive lower bounds in each of the quantities, $\Delta x$, $\Delta y$, $\Delta p_x$ and $\Delta p_y$.

At small $T$, we calculate the radial moments using saddle-point approximation \cite{supp} and find that in the $T\to 0$ limit $\ev{r}  = r_c$  and   $\ev{r^n} =\ev{r}^n\ \forall n\geq0.$  Thus, certainly the particle remains on $r=r_c$ as $ T\to0,$ having a residual energy $\ev{E}=E_c.$  Fluctuations in $r$ and $E$ vanish in this limit, but the uncertainty relation \eqref{uncertaintycons} holds  with the equality valid as $T\to0$.

\subsection{Power Law Potentials \label{sec:IIA}}
\begin{figure}[t]
    \centering
    \includegraphics[width=\linewidth]{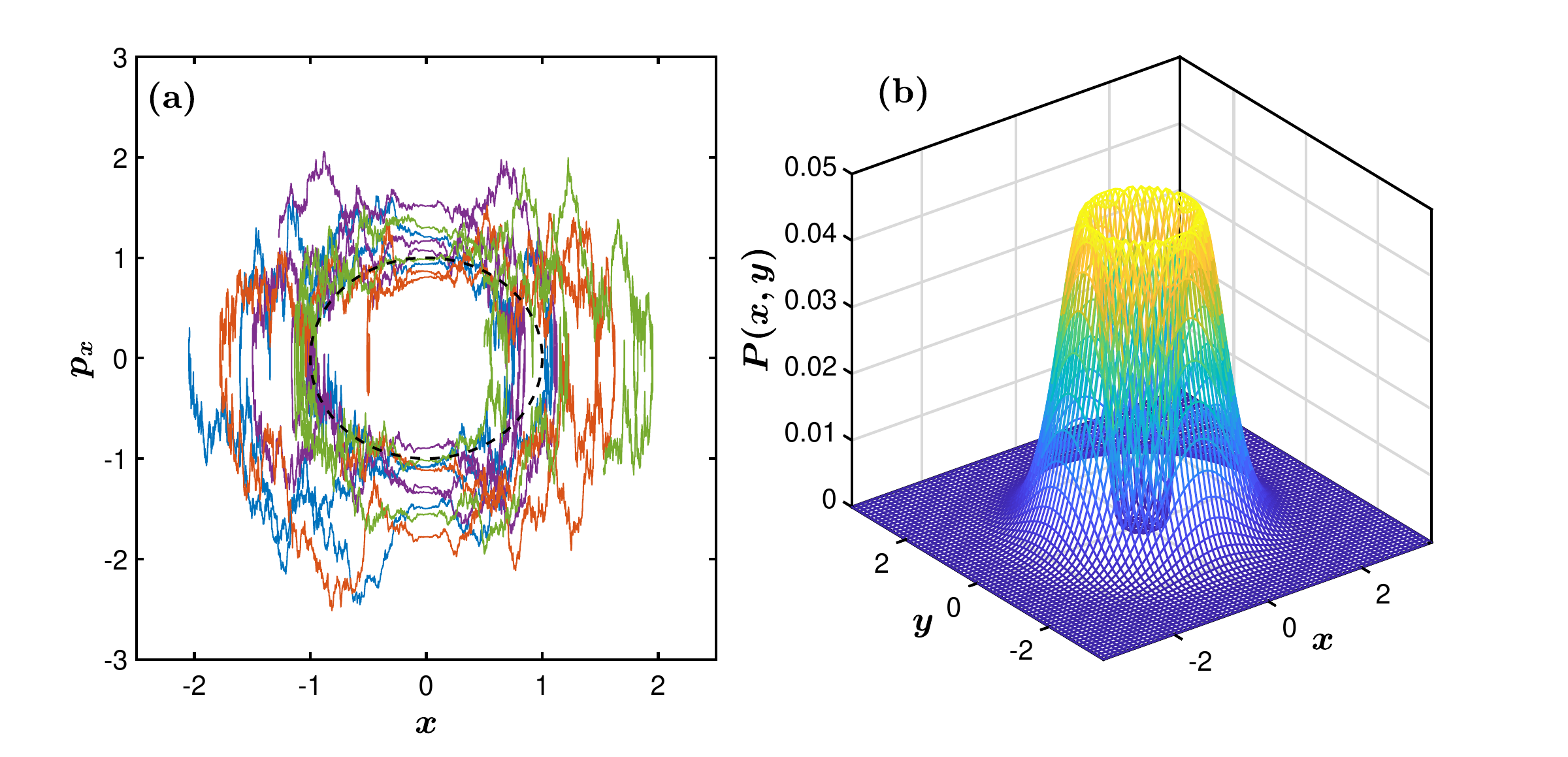}
\vspace*{-.7cm}

    \caption{(a) Four sample paths, (b) steady state distribution $P(x,y)$, for $V(r)=r^2/2$ with conserved  $\ell=1$  ($\gamma=2$, $T=0.5$). Dashed line in (a): limiting trajectory  as $T\to0.$}\vspace*{-.4cm}
    
    \label{fig:distribution_sho}
\end{figure}

As an example, let us  consider the potentials
\begin{equation}
    V(r)=\frac{kr^\alpha}{\alpha};\quad k,\alpha>0,
\end{equation}
which yield the integral in \eqref{conditioncons} finite. 
In the limit of vanishing noise, ($T\to0$), the particle is constrained to move in a circular trajectory in the $x$-$y$ plane having radius $r_c$ and energy $E_c,$ given by
\begin{equation}
    r_c=\qty(\dfrac{\ell^2}{k})^{\frac{1}{\alpha+2}}\qcomma E_c= k \qty(\frac{2+\alpha}{2\alpha})r_c^\alpha.
    \label{eq:rc}
\end{equation}
The sample paths and steady state distribution for $\alpha=2$ (simple harmonic oscillator potential) have been shown in Fig. \ref{fig:distribution_sho}. There, we find $\Delta x=\frac{1}{\sqrt{2}}\sqrt{\frac{T}{k}+\frac{\abs{\ell}}{\sqrt{k}}}$ and $\Delta p_x=\frac{1}{\sqrt{2}}\sqrt{T+\abs{\ell}\sqrt{k}}$. Both of these quantities are increasing functions of $T$ and have a lower bound as $T\to0$ given by $\min(\Delta x)=\frac{r_c}{\sqrt{2}}$ and $\min(\Delta p_x)=\frac{\abs{\ell}}{r_c\sqrt{2}}$, both of which are dependent on $k$. However, the lower bound in their product $\min(\Delta x\Delta p_x)=\min(\Delta x)\min(\Delta p_x)=\frac{\abs{\ell}}{2}$ is independent of $k$.

\begin{figure}[t]
    \centering
    \includegraphics[width=\linewidth]{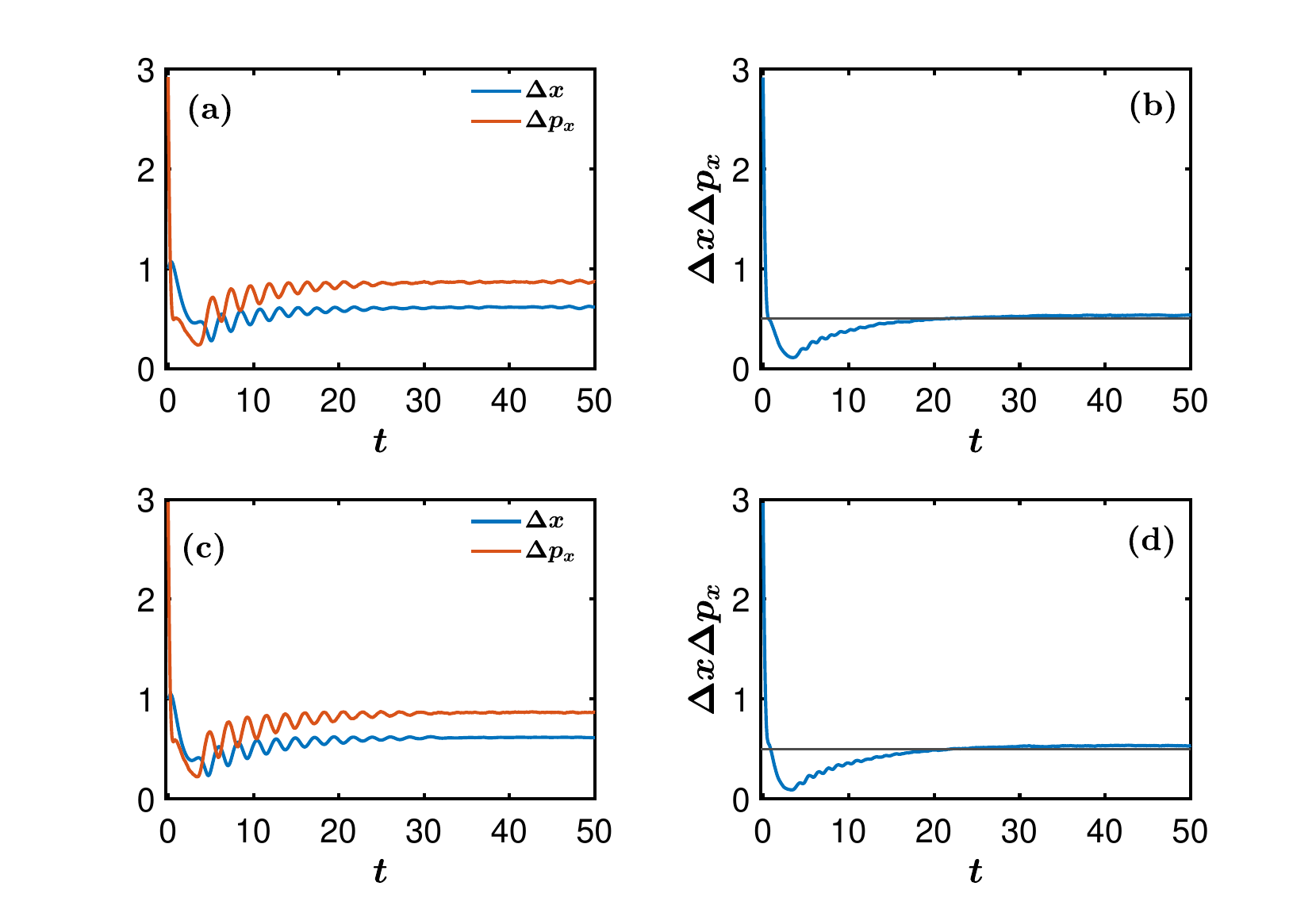}
    \caption{Time evolution of $\Delta x$, $\Delta p_x$ and their product $\Delta x\Delta p_x$ for $V(r)=kr^\alpha/\alpha$ with $\gamma=4$, $k=2$ and conserved $\ell=1$ at $T=10^{-1}$. (a) and (b) show this for $\alpha=2$. (c) and (d) show this for $\alpha=2.1$. The lower bound, $\abs{\ell}/2$ of $\Delta x\Delta p_x$, is shown in a black solid line in (b) and (d). Data is averaged over 5000 samples.  The stochastic differential equations \eqref{eq:2dLangevin} were integrated using the second order technique discussed in \cite{Mannella1989}.}
    \label{fig:cons_time}
\end{figure}

With the dynamics specified, we examine how $\Delta x \Delta p_x$ evolves to its steady state value, which is always larger than or equal to the lower bound $\abs{\ell}/2$ imposed by the uncertainty relation \eqref{uncertaintycons}. Figure~\ref{fig:cons_time} shows the time evolution of $\Delta x$, $\Delta p_x$, and their product $\Delta x\Delta p_x$ for central potentials with $\alpha=2$ and $\alpha=2.1$. By selecting an appropriate initial distribution of $(x,y)$ and $(p_x,p_y)$ at a fixed angular momentum $\ell$, one can make $\Delta x$ and $\Delta p_x$ arbitrarily large or small at $t=0$. Nevertheless, numerical simulations of the dynamics \eqref{eq:2dLangevin} reveal that as the system relaxes to steady state, $\Delta x\Delta p_x$ always remains greater than or equal to the lower bound, which depends solely on $\abs{\ell}$. This confirms that the bound is an intrinsic steady state property of the system. Interestingly, for both $\alpha=2$ and $\alpha=2.1$, $\Delta x$ and $\Delta p_x$ display out-of-phase oscillations as the steady state is approached—an increase in $\Delta x$ coincides with a decrease in $\Delta p_x$—closely resembling the behavior observed in quantum systems.

\section{\label{sec:nonconsL}Breaking conservation of angular momentum}  
It might appear that  the robustness  of the  uncertainty relation  owes to the conservation of angular  momentum,  which  forces  the system to be  on  a  closed curve   when  noise is absent. This is, however,  not true. In fact, we  show  below that  the    position-momenta   uncertainty relation \eqref{uncertaintycons} carries forward  to  systems where the angular momentum is not conserved. In the non-conserved ensemble, a generalized chemical potential  controls the average  angular momentum $p_\phi.$  The dynamics of such a system is governed by 
\begin{equation}
\begin{aligned}
    &\dot{r}=p_r\qcomma & &\dot{p}_r=-\pdv{V}{r}+\dfrac{p_\phi^2}{r^3}\\
    &\dot{\phi}=\dfrac{p_\phi}{r^2}-\mu\qcomma & &\dot{p}_\phi=\mu r^2\gamma-\gamma p_\phi+r\sqrt{\gamma T}\xi(t). \label{eq:Lang_2d_noncons}
\end{aligned}
\end{equation}
The dynamics is chosen suitably (see Appendix \ref{app:FP} for details), so that the steady state of the corresponding Fokker-Planck equation,
\begin{equation}
    \pdv{P}{t}=[\widetilde{H},P]+\pdv{p_\phi}((\gamma p_\phi-\mu r^2\gamma)P+r^2\gamma T\pdv{P}{p_\phi}), \label{eq:FPE_noncons}
\end{equation}
with $\widetilde{H}=H-\mu p_\phi$, is an equilibrium distribution  $P(r, p_r, p_\phi)= \frac1{\mathcal{Z}(\beta,\mu)} \exp (-\beta (H -\mu p_\phi)),$ where  $\mathcal{Z}(\beta,\mu)$, the partition function reads as
\begin{equation}
    \mathcal{Z}(\beta,\mu)=\dfrac{4\pi^2}{\beta}\int_0^{\infty}\dd{r} r\exp(-\beta\qty(V(r)-\dfrac{\mu^2r^2}{2})). \label{conditionnoncons}
\end{equation} 
As in the conserved case, $\mathcal{Z}(\beta,\mu)$ must be finite for $P(r,p_r,p_\phi)$ to be a valid PDF. 

A system described by a non-zero $\mu$ can have a steady state only if the radial potential $V(r)$ grows asymptotically as $r^2$ or faster. Such potentials  lead  us to the following relation between moments \cite{supp}
\begin{equation}
\begin{aligned}
    \ev{p_\phi}&=\mu\ev{r^2}= \ev{\frac{p_\phi}{r^2}}\ev{r^2}, \\
    \ev{\dfrac{p_\phi^2}{2r^2}}&=\dfrac{T}{2}+\dfrac{\mu^2\ev{r^2}}{2}=\dfrac{T}{2}+\dfrac{\ev{p_\phi}^2}{2\ev{r^2}}.
\end{aligned}
\end{equation}
Provided $\ev{r^2}$ is finite, we can always choose a $\mu$ which allows $p_\phi$ to fluctuate about a desired mean $\ev{p_\phi}$. 

As before, $\ev{x^n}=\ev{y^n}, \ev{p_x^n}=\ev{p_y^n} \forall n$, all vanishing for odd $n$, and
\begin{align}
    \ev{x^2}=\dfrac{\ev{r^2}}{2}\qcomma \ev{p_x^2}=\dfrac{T}{2}+\ev{\dfrac{p_\phi^2}{2r^2}}.
\end{align} 
Therefore,
\begin{equation}
    \Delta x\Delta p_x=\Delta y\Delta p_y=
    \frac12\sqrt{2T \ev{r^2} +\ev{p_\phi}^2}.
\end{equation}
 As earlier, $T\ev{r^2}\geq0$, using which, we conclude $\sqrt{2T \ev{r^2} +\ev{p_\phi}^2}\geq\sqrt{\ev{p_\phi}^2}$ and we end up with the uncertainty relations
\begin{equation}
    \Delta x\Delta p_x=\Delta y\Delta p_y\geq\dfrac{\abs{\ev{p_\phi}}}{2},\label{uncertaintynoncons}
\end{equation}
where, akin to the conserved angular momentum ensemble, the equality holds in the limiting sense as $T\to0$.

\subsection{Power Law Potentials:\boldmath\texorpdfstring{$\alpha=2$ vs $\alpha\neq2$}{alpha=2 vs alpha neq 2}}
In this  non-conserved angular momentum ensemble, particles in a radial  potential $V(r)=kr^\alpha/\alpha$  can have  a steady state when $\alpha\geq2,$ as required by \eqref{conditionnoncons}. For any $\alpha>2$,  the mean angular momentum of the system can be  fixed at  any desired value  $\ev{p_\phi}=\ell$ by suitably  choosing   the  corresponding chemical potential  $\mu$ in the  range  $\mu\in\qty(-\infty, \infty).$ Then, in the $T\to0$ limit,  $\ev{r}$  approaches the  value $r_c$  given   in  Eq.  \eqref{eq:rc}  when   $\mu$ is set such that $\ev{p_\phi}=\ell.$   At low  temperature, particles  make noisy motion about this  circle $r=r_c$ as  shown in Fig. \ref{fig:noncons_samplepaths} (b). Like in the conserved ensemble, here too,  both $\Delta r$ and $\Delta E$, along with $\Delta p_\phi$, vanish in the  $T\to0$ limit  \cite{supp}, but  the uncertainty relation  \eqref{uncertaintynoncons} holds.

Nonexistence of a steady state  for particles  in a central potential with $\alpha<2$    brings   the harmonic case,  $\alpha=2$,   into focus.  Being  on the boundary of  having or not having a steady state,  $\alpha=2$   is   quite special.  Here  the  chemical potential   has a finite  bound  $\mu\in\qty(-\sqrt{k},\sqrt{k});$  one can get any desired  value $\ev{p_\phi}=\ell$  by tuning $\mu$ in this finite range. The consequences of this are manifest in the behaviour of the system as follows.    When $T\to0,$   the average 
 value  of  $\ev{r} \to \frac{\sqrt{\pi}}{2}r_c$ ( smaller than $r_c= (\ell^2/k)^{1/4}$),  and  there  are finite fluctuations of  particle around the circle  $\ev{r} = \frac{\sqrt{\pi}}{2}r_c,$ i.e. $\Delta r(T\to0)=r_c\sqrt{\frac{4}{\pi}-1}\neq0$, even when  $T\to0.$  The  trajectory of particles  at small temperature is  shown in Fig. \ref{fig:noncons_samplepaths} (a) -  trajectories for $\alpha=2$ are strikingly different from $\alpha=2.1.$ Furthermore, $\Delta E(T\to0)=\sqrt{k}\abs{\ev{p_\phi}}$ and hence, the specific heat $C_V=\frac{\Delta E^2}{T^2}$ diverges as $T\to0$. Unlike, $\alpha>2$, here we also have $\Delta p_\phi(T\to0)=\abs{\ev{p_\phi}}$.  Explicit details for the simple harmonic oscillator in the non-conserved angular momentum bath, along with its differences from $\alpha>2$ systems, can be found in \cite{supp}.

\begin{figure}[t]
    \centering
    \includegraphics[width=\linewidth]{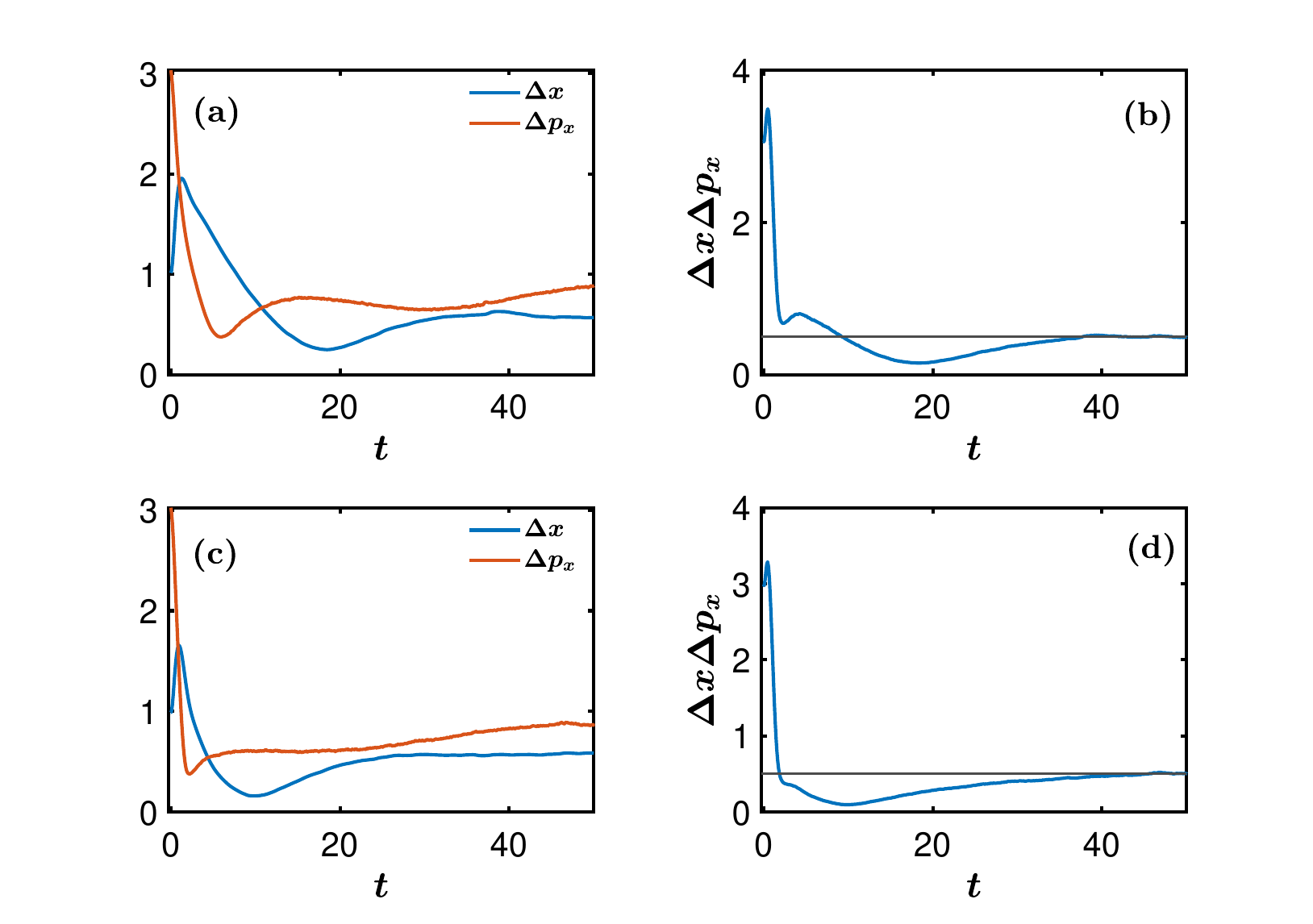}
    \caption{Time evolution of $\Delta x$, $\Delta p_x$ and their product $\Delta x\Delta p_x$ for $V(r)=kr^\alpha/\alpha$ with $\gamma=4$, $k=2$ and fluctuating $p_\phi$ with $\ev{p_\phi}=1$ at $T=10^{-1}$. (a) and (b) show this for $\alpha=2$. (c) and (d) show this for $\alpha=2.1$. The lower bound, $\abs{\ev{p_\phi}}/2$ of $\Delta x\Delta p_x$, is shown in a black solid line in (b) and (d). Data is averaged over 5000 samples.  The stochastic differential equations \eqref{eq:Lang_2d_noncons} were integrated using the second order technique discussed in \cite{Mannella1989}.}
    \label{fig:noncons_time}
\end{figure}
As in the conserved angular momentum ensemble, we numerically investigate how the uncertainties relax following   Eq. \eqref{eq:Lang_2d_noncons} to their steady state values in the non-conserved case. Figure~\ref{fig:noncons_time} shows the time evolution of $\Delta x$, $\Delta p_x$, and their product $\Delta x\Delta p_x$ for central potentials with $\alpha=2$ and $\alpha=2.1$. The parameter $\mu$ is chosen such that the steady-state angular momentum satisfies $\langle p_\phi \rangle = 1$ at $T = 10^{-1}$. As before, the initial distribution of $(x,y,p_x,p_y)$ is chosen to make $\Delta x$ and $\Delta p_x$ arbitrarily large at $t=0$. Despite these initial conditions, $\Delta x\Delta p_x$ in both cases approaches a steady state value that exceeds the lower bound $\abs{\langle p_\phi \rangle}/2$, consistent with the uncertainty relation \eqref{uncertaintynoncons}. In contrast to the conserved ensemble, however, the approach to steady state is monotonic, and no out-of-phase oscillations between $\Delta x$ and $\Delta p_x$ are observed.

\subsection{\label{sec:discussion}Discussion: Physical Nature of the Baths}

 At the outset, the presented construction of the conserved angular momentum bath (in Sec. \ref{sec:consL})   or the non-conserved angular momentum bath (this section), valid for any central potential   may appear rather artificial. However, upon further consideration, one may find it is not the case. The dynamics has been constructed staying true to the concept of the generalized Gibbs ensemble \cite{Gibbs2010}, where in steady state, the weight of a configuration is given by $\exp(-\beta H)\exp(\beta\sum_{i}\lambda_i O_i)$ where $\lambda_i$s are the Lagrange multipliers coupling to the corresponding non-conserved observables $O_i$s such that they can be tuned, i.e., $\ev{O_i}$ can be set to any desired value.

The conserved angular momentum bath physically describes a rich class of physical systems, namely, limit cycle oscillators under temperature and has been discussed in detail in \cite{singh2025limitcycle}.

In the non-conserved dynamics \eqref{eq:Lang_2d_noncons}, only the $\dot p_\phi$ equation, i.e., the torque equation of the Hamiltonian dynamics was modified. At first, it may appear strange that an explicit noise source was added only to the torque equation and not the equation of radial force, i.e., the $\dot p_r$ equation, as is the case in usual Langevin dynamics. We can, in fact, use an alternative dynamics with independent noise sources for both $\dot p_r$ and $\dot p_\phi$ equations given by
\begin{equation}
\begin{aligned}
    &\dot{r}=p_r\qcomma & &\dot{p}_r=-\pdv{V}{r}+\dfrac{p_\phi^2}{r^3}-\gamma_2p_r+\sqrt{\gamma_2T}\xi_2(t)\\
    &\dot{\phi}=\dfrac{p_\phi}{r^2}-\mu\qcomma & &\dot{p}_\phi=\mu r^2\gamma_1-\gamma_1 p_\phi+r\sqrt{\gamma_1 T}\xi_1(t), \label{eq:alternativeLang_2d_noncons}
\end{aligned}
\end{equation}
with $\ev{\xi_i(t)}=0$ and $\ev{\xi_i(t)\xi_j(t')}=2\delta_{ij}\delta(t-t')$ where $i,j=1,2$. The Fokker-Planck equation for the dynamics \eqref{eq:alternativeLang_2d_noncons} reads
\begin{equation}
\begin{aligned}
    \pdv{P}{t}=[&\widetilde{H},P]+\pdv{p_r}(\gamma_2p_rP+\gamma_2T\pdv{P}{p_r})\\
    &+\pdv{p_\phi}((\gamma_1 p_\phi-\mu r^2\gamma_1)P+r^2\gamma_1 T\pdv{P}{p_\phi})
\end{aligned}
\end{equation}
with $\widetilde H=H-\mu p_\phi$. The above Fokker-Planck equation also admits the same steady state solution, $P\sim \exp(-\beta\widetilde H)=\exp(-\beta H)\exp(\beta\mu p_\phi)$. Thus, adding an additional noise source to the $\dot p_r$ equation is redundant and modifying the $\dot p_\phi$ equation is enough. The noise explicitly supplied to the $\dot p_\phi$ equation is seen by all other coordinates as the stochastic differential equations are coupled. 

In further discussion on the nature of the non-conserved bath, we have referred to $\mu$ as a generalized chemical potential, and we should be able to associate $\mu$, the defined Lagrange multiplier to a physical quantity just like it is possible to do so with other Lagrange multipliers. For instance, $\beta$ which couples to energy is inverse temperature, Lagrange multiplier coupling to the number of particles  is `chemical potential' while the one coupling to magnetization is magnetic field, and so on. 

To understand the physical nature of $\mu$, we fall back on our constructed dynamics \eqref{eq:Lang_2d_noncons}, particularly the $\dot\phi=\frac{p_\phi}{r^2}-\mu$ and $\dot p_\phi=\gamma(\mu r^2-p_\phi)+r\sqrt{\gamma T}\xi(t)$ equations. Moreover, we also use the steady state relation $\mu=\ev{\frac{p_\phi}{r^2}}=\frac{\ev{p_\phi}}{\ev{r^2}}$. It is clear that by choosing a $\mu$, we actually end up setting the average angular velocity such that $\ev{\dot\phi}=0$ in the steady state. Moreover, $\mu$ appears as $\gamma(\mu r^2-p_\phi)$, an external torque term, in the $\dot p_\phi$ equation, forcing $\ev{p_\phi}=\mu\ev{r^2}$ in steady state, in turn setting $\ev{p_\phi}$ to the chosen value. Thus, $\mu$ is nothing but angular velocity which couples to the angular momentum. Brownian gyrators \cite{Filliger2007} can be a possible way of providing the required fluctuating torque to the system and hence, realizing the non-conserved bath experimentally.

\section{\label{sec:noncentral}Non-central Potentials}
So far, we have restricted ourselves to 2D central potentials. We extend our analysis further to 2D $\phi$-dependent potentials. Consider the Hamiltonian
\begin{equation}
    H=\frac{p_r^2}{2}+\frac{p_\phi^2}{2r^2}+V(r,\phi),
\end{equation}
with $V(r,\phi+2\pi)=V(r,\phi)$. Obviously, $\phi$ is no more cyclic and we directly work in the non-conserved angular momentum ensemble. The dynamics is same as \eqref{eq:Lang_2d_noncons} but for an additional $-\pdv*{V}{\phi}$ term in the $\dot p_\phi$ equation with the same Fokker-Planck equation as Eq. \eqref{eq:FPE_noncons}. Now, the equilibrium distribution reads $P(r,p_r,\phi,p_\phi)=\frac{1}{\mathcal{Z}(\beta,\mu)}\exp(-\beta(H-\mu p_\phi))$ with the partition function
\begin{equation}
    \mathcal{Z}(\beta,\mu)=\frac{2\pi}{\beta}\int_{0}^{2\pi}\dd{\phi}\int_{0}^{\infty}\dd{r} r\ e^{-\beta\qty(V(r,\phi)-\frac{\mu^2r^2}{2})},
\end{equation}
which should be finite for $P(r,p_r,\phi,p_\phi)$ to be a valid PDF. Due to $\phi$-dependence, $\ev{x^n}\neq\ev{y^n}$ and similarly, $\ev{p_x^n}\neq\ev{p_y^n}$ in general. However, using the relations between moments \cite{supp}, we have
\begin{align}
    \Delta p_x^2=T+\frac{\ev{p_\phi}^2}{\ev{r^2}^2}\Delta y^2;~~ \Delta p_y^2=T+\frac{\ev{p_\phi}^2}{\ev{r^2}^2}\Delta x^2.
\end{align}
Provided $\ev{r^2}$ exists, a $\mu$ can be chosen such that $p_\phi$ fluctuates about a given $\ev{p_\phi}$ in steady state and we obtain the following uncertainty relations
\begin{equation}
\begin{aligned}
    &\Delta x \Delta p_x\ge c_{xx} {\abs{\ev {p_\phi}}};~~ \Delta y \Delta p_y \ge c_{yy} {\abs{\ev {p_\phi}}}\\
    &\Delta x \Delta p_y\ge c_{xy} {\abs{\ev {p_\phi}}};~~ \Delta y \Delta p_x \ge c_{yx} {\abs{\ev {p_\phi}}}
    \label{eq:gen_relation}
\end{aligned}
\end{equation}
where $c_{xx}=c_{yy}=c= \min\qty(\frac{\Delta x\Delta y}{\ev{r^2}})\in[0,1/2]$, $c_{xy}=\min\qty(\frac{\Delta x^2}{\ev{r^2}})\in[0,1]$ and $c_{yx}=\min\qty(\frac{\Delta y^2}{\ev{r^2}})\in[0,1]$ are dimensionless parameters. The equalities hold when $c_{ij}=\qty(\frac{\Delta q_i\Delta q_{3-j}}{\ev{r^2}})(T\to0)$ where $q_1=x$ and $q_2=y$  (see Appendix \ref{app:unc_noncentral} for details).

Owing to the $\phi$-dependence of the potential, we may have a trivial lower bound in the uncertainty between conjugates, i.e., $c=0$ but one of $c_{xy}$ or $c_{yx}$ may still be non-zero. Therefore, depending on the form of the potential, we may have a trivial uncertainty relation between conjugates but a non-trivial lower bound for non-conjugates. Physically, this can be understood from the symmetry or equivalently, asymmetry of $V(r,\phi)$. As one would expect, the probability of finding the particle will be more around the minima of the potential, being finite around these regions and zero if away as $T\to0$. Now, if the minima are placed, say along the $x$-axis, then, $\Delta y$ and hence, $\Delta p_x$ vanish as $T\to0$. However, we shall always have a non-zero minimum in $\Delta x$ for some $T$. Thus, $c=0=c_{yx}$ but $c_{xy}\neq0$. If all the minima are symmetrically placed wrt both $x$ and $y$ then we shall have $\ev{x^2}=\ev{y^2}=\frac{1}{2}\ev{r^2}$ and $c=c_{xy}=c_{yx}=\frac{1}{2}$. If the minima are asymmetric wrt both $x$ and $y$, then we may even have $c\neq c_{xy}\neq c_{yx}$.

\section{\label{sec:higherD}Higher Dimensions}
The   uncertainty  relations  derived in Eq. \eqref{eq:gen_relation} can be generalized to higher dimensions. In a $3$D central potential, the problem of the conserved angular momentum ensemble can be easily reduced to the $2$D case considered in the article by a simple rotation of coordinates. As $p_\phi=\ell$ is fixed, we can always rotate our coordinates such that $p_\phi$ is along the $z$-direction, making the problem effectively two dimensional. This is equivalent to choosing the initial conditions $(\theta(t=0),p_\theta(t=0))=(\pi/2,0)$ which would restrict the motion in the $x$-$y$ plane. Thus, the uncertainty relation \eqref{uncertaintycons} holds and carries forward to the relation \eqref{uncertaintynoncons} in the non-conserved ensemble. We explicitly show in \cite{supp} that $\Delta x\Delta p_x=\Delta y\Delta p_y\geq\abs{\ev{p_\phi}}/2$ with $\Delta z\Delta p_z\geq0$ for $3$D central potentials. We believe this notion can be extended to a general potential $V(r,\theta,\phi)$ in $3$D. We therefore propose for a  system of  several   conjugate variables (co-ordinates $\{q_i\}$ and momenta $\{p_i\})$  with finite  mean angular momentum,   
\begin{equation}
\Delta q_i \Delta p_j \ge c_{ij}  \abs{\ev{p_\phi}};~~~~~ 0\le c_{ij} \le 1-\frac12 \delta_{ij}.
\end{equation}

\section{\label{sec:conc}Conclusion}
In conclusion, we  show  that whenever a particle is constrained to have $\ev{p_\phi}\neq0$ at all temperatures, in  equilibrium,  its position $q_i$ and momenta $p_i$ would exhibit an uncertainty  relation $\Delta q_i \Delta p_j \ge c_{ij}\abs{\ev{p_\phi}}$  with  at least one of  $c_{ij}\ne0,$ i.e., the  product of the standard deviations  of these  observables  is bounded from below by a  finite  positive constant.  For particles in 2D central potentials, $c_{ij}=\frac{1}{2}$, $i,j=1,2$. 
When  the  potential is $\phi$-dependent, the uncertainty relation  is modified to Eq. \eqref{eq:gen_relation}. 

The emergence of an uncertainty relation can be traced to the system possessing non-zero average angular momentum  $\ev{p_\phi} \ne 0$ in the  $T\to 0$  limit.  This forces   the particle  to fluctuate about a circular  trajectory in the $x$-$y$ plane, resulting  in  non-trivial lower bounds on the fluctuations of $x$, $p_x=\dot x$, $y$ and $p_y=\dot y$. Conventional Langevin baths regulate the system's average energy through $\beta$ (a Lagrange multiplier),  while allowing $p_\phi$ to take all possible values with equal-a priori-probability,   forcing  $\ev{p_\phi} =0$  at any temperature.  To have a non-zero $\ev{p_\phi}$, the thermal bath must include an additional chemical potential $\mu$  that either couples to  $p_\phi$
or ensures angular momentum conservation, allowing $p_\phi$ to retain its initial value while permitting total energy fluctuations. 
In this work, we explicitly construct thermal baths that enable both energy and angular momentum to fluctuate around chosen mean values, governed by the generalized chemical potentials $\beta$ and $\mu,$
 respectively. These baths drive the system toward an equilibrium state characterized by a Boltzmann distribution.

We believe the  theoretical predictions  of the  model can be verified in  systems  where  particles move in a  noisy trajectory around a closed curve, even in the limit $T\to0$. One such system is  an active  Brownian particle   in a  harmonic trap  in two dimensions \cite{Basu2019, Dauchot2019}; for very high motility, they  move  stochastically  around a  circle, far away from the  minimum of the potential.  Another example is the  chiral active particles, which  naturally form     noisy  trajectories around   a circle \cite{Kmmel2013, tenHagen2014}. 


\appendix
\section{\label{app:FP}Fokker-Planck Equation}
For a set of $N$-variable Langevin equations of the form 
\begin{equation}
    \dot{\zeta}_i=h_i(\zeta,t)+\sum_{j}g_{ij}(\zeta,t)\xi_j(t),
\end{equation}
where $\xi_i(t)$ are Gaussian white noise with $\ev{\xi_i(t)}=0$ and $\ev{\xi_i(t)\xi_j(t')}=2\delta_{ij}\delta(t-t')$, the Fokker-Planck equation for the probability density $P(\zeta,t)$ is given by
\begin{equation}
    \pdv{P}{t}=-\sum_{i}\pdv{\zeta_i}(D_iP)+\sum_{i,j}\frac{\partial^2}{\partial \zeta_i\partial \zeta_j}(D_{ij}P),
\end{equation}
where $\displaystyle D_i=h_i+\sum_{k,j}\pdv{g_{ij}}{\zeta_k}g_{kj}$ and $D_{ij}=\displaystyle\sum_{k}g_{ik}g_{jk}$ are the so-called drift and diffusion coefficients respectively \cite{Risken1996}.

For a noisy dynamics derived from the dynamics generated by a Hamiltonian $H$, as is our case, with the equations
\begin{equation}
    \dot{x}_i=\pdv{H}{p_i}\qcomma \dot{p}_i=-\pdv{H}{x_i}-f_i(x,p)+g_i(x,p)\xi_i(t),
\end{equation}
with $\ev{\xi_i(t)}=0$ and $\ev{\xi_i(t)\xi_j(t')}=2\delta_{ij}\delta(t-t')$, the non-trivial drift and diffusion coefficients are
\begin{equation}
\begin{aligned}
    D_{x_i}&=\pdv{H}{p_i}\qcomma\\
    D_{p_i}&=-\pdv{H}{x_i}-f_{i}+g_{i}\pdv{g_{i}}{p_i} \qand\\
    D_{p_i p_i}&=g_{i}^2,
\end{aligned}
\end{equation}
using which the Fokker-Planck equation for probability density $P(x,p,t)$ can be written as 
\begin{equation}
    \pdv{P}{t}=[H,P]+\sum_{i}\pdv{p_i}(f_i P-g_{i}\pdv{g_{i}}{p_i}P+\pdv{p_i}(g_i^2 P)),
\end{equation}
where $[A,B]=\displaystyle\sum_{i}\qty(\pdv{A}{x_i}\pdv{B}{p_i}-\pdv{A}{p_i}\pdv{B}{x_i})$ is the Poisson bracket of $A$ and $B$. For the special case when $g_{i}$ is independent of $p_i$, the Fokker-Planck equation becomes
\begin{equation}
    \pdv{P}{t}=[H,P]+\mathcal{L}P=[H,P]+\sum_{i}\pdv{p_i}(f_i P+g_i^2\pdv{P}{p_i}).
\end{equation}

Corresponding to Langevin equations 
\begin{equation}
    \dot{r}=p_r,\ \dot{p}_r=-V'(r)+\frac{\ell^2}{r^3}-\gamma p_r+\sqrt{\gamma T}\xi(t),\ \dot{\phi}=\frac{\ell}{r^2}, \label{eq:radialLangevin}
\end{equation}
the Fokker-Planck is given by
\begin{equation}
    \pdv{P}{t}=[H,P]+\pdv{p_r}(\gamma p_rP+\gamma T\pdv{P}{p_r}).
\end{equation}
In steady state, $\partial_t P=0$. We then use the ansatz $P=P(H)$, and noting that $[H,P(H)]=0$ and $\pdv{H}{p_r}=p_r$, we have
\begin{equation}
\begin{aligned}
    \pdv{p_r}\biggl(\gamma p_r P+&\gamma T\pdv{P}{H}\pdv{H}{p_r}\biggr)\\
    &=\pdv{p_r}(\gamma p_r\qty( P+ T\pdv{P}{H}))=0.
\end{aligned}
\end{equation}
Thus, $P(H)\sim \exp(-\beta H)$ where $\beta\equiv T^{-1}$ solves the above equation. We demand that the probability density integrated over all space converges and hence, the partition function
\begin{equation}
\begin{aligned}
    \mathcal{Z}_\ell(\beta)&=\int_{0}^{2\pi}\dd{\phi}\int_{-\infty}^{\infty}\dd{p_r}\int_{0}^{\infty}\dd{r}P(r,p_r,\ell)\\
    &=\int_{0}^{2\pi}\dd{\phi}\int_{-\infty}^{\infty}\dd{p_r}\int_{0}^{\infty}\dd{r}\exp(-\beta H)=\mathrm{finite}.
\end{aligned}
\end{equation}
Integrating $\phi$ and $p_r$, the condition reduces to
\begin{equation}
    \Z_\ell(\beta)=2\pi\sqrt{\dfrac{2\pi}{\beta}}\int_{0}^{\infty}\dd{r}\ e^{-\beta\qty(\frac{\ell^2}{2r^2}+V(r))}=\mathrm{finite}.
\end{equation}
If the above holds, $P(r,p_r,\ell)=\frac{e^{-\beta H}}{\mathcal{Z}_\ell(\beta)}$ is the steady state solution.

We defined a dynamics which allowed $p_\phi$ to fluctuate about a desired mean $\ev{p_\phi}$. To do so, we define an auxiliary Hamiltonian,
\begin{equation}
    \widetilde{H}=H(r,p_r,\phi,p_\phi)-\mu p_\phi=\frac{p_r^2}{2}+\frac{p_\phi^2}{2r^2}+V(r,\phi)-\mu p_\phi,
\end{equation}
and demand that the steady state weight is $\exp(-\beta \widetilde H)=\exp(-\beta H)\exp(\beta\mu p_\phi)$. We write down the Hamiltonian equations of motion for $\widetilde H$,
\begin{equation}
    \dot r=\pdv{\widetilde H}{p_r},\ \dot p_r=-\pdv{\widetilde H}{r},\  \dot\phi=\pdv{H}{p_\phi}-\mu,\ \dot p_\phi=-\pdv{\widetilde H}{\phi}.
\end{equation}
We need a fluctuating angular momentum, so we modify the $\dot p_\phi$ equation. $\dot p_\phi$ has torque dimension, so we multiply the Langevin force with $r$ to create the simplest torque dimension object using the Langevin force. Obviously, we need to add a dissipation and an additional forcing such that $p_\phi$ fluctuates about $\ev{p_\phi}$ in steady state. Let this be encoded in a now undetermined forcing $f$. Thus, the $\dot p_\phi$ equation is modified as 
\begin{equation}
    \dot p_\phi=f+r\sqrt{\gamma T}\xi(t).
\end{equation}
The Fokker-Planck equation in steady state reads
\begin{equation}
    [\widetilde H,P]+\pdv{p_\phi}(-fP+r^2\gamma T\pdv{P}{p_\phi})=0.
\end{equation}
We use the ansatz $P=P\qty(\widetilde H)$. The Poisson bracket vanishes and we get
\begin{equation}
\begin{aligned}
    &\pdv{p_\phi}(-fP+r^2\gamma T\pdv{P}{\widetilde H}\pdv{\widetilde H}{p_\phi})\\
    &=\pdv{p_\phi}(-fP+r^2\gamma T\pdv{P}{\widetilde H}\qty(\dfrac{p_\phi}{r^2}-\mu))\\
    &=\pdv{p_\phi}(-fP+T\pdv{P}{\widetilde H}\qty(\gamma p_\phi-\mu r^2\gamma))=0.
\end{aligned}
\end{equation}
We make a choice
\begin{equation}
    f=\mu r^2\gamma-\gamma p_\phi,
\end{equation}
using which
\begin{equation}
    \pdv{p_\phi}((\gamma p_\phi-\mu r^2\gamma)\qty(P+T\pdv{P}{\widetilde H}))=0,
\end{equation}
has the obvious solution $P(\widetilde H)\sim e^{-\beta\widetilde H}=e^{-\beta H}e^{\beta\mu p_\phi}$. We demand that the probability density integrated over all space to be finite. That is, the partition function,
\begin{equation}
\begin{aligned}
    \Z(\beta,\mu)&=\int_{-\infty}^{\infty}\dd{p_r}\int_{-\infty}^{\infty}\dd{p_\phi}\int_{0}^{2\pi}\dd{\phi}\int_{0}^{\infty}\dd{r}e^{-\beta H}e^{\beta\mu p_\phi}\\
    &=\mathrm{finite}.
\end{aligned}
\end{equation}
Integrating over $p_r$ and $p_\phi$, the condition reduces to 
\begin{equation}
\begin{aligned}
    \Z(\beta,\mu)&=\frac{2\pi}{\beta}\int_{0}^{2\pi}\dd{\phi}\int_{0}^{\infty}\dd{r} r\ e^{-\beta\qty(V(r,\phi)-\frac{\mu^2r^2}{2})}\\
    &=\mathrm{finite}.
\end{aligned}
\end{equation}
If the above holds, $P(r,p_r,\phi,p_\phi)=\frac{e^{-\beta H}e^{\beta\mu p_\phi}}{\mathcal{Z}(\beta,\mu)}$ is the steady state solution.

A simple exercise would be to extend this idea of demanding a steady state to define the dynamics in $3$D to have $\exp(-\beta\widetilde H)$ as the steady state where
\begin{equation}
\begin{aligned}
    \widetilde H&=H(r,p_r,\theta,p_\theta,\phi,p_\phi)-\mu p_\phi\\
    &=\frac{p_r^2}{2}+\frac{p_\theta^2}{2r^2}+\frac{p_\phi^2}{2r^2\sin^2\theta}+V(r,\theta,\phi)-\mu p_\phi.
\end{aligned}
\end{equation}
In $3$D, dynamics \eqref{eq:sm_3D_noncons} produces $P\qty(\widetilde H)\sim e^{-\beta\widetilde H}=e^{-\beta H}e^{\beta\mu p_\phi}$
\begin{equation}
\begin{aligned}
    &\dot r=p_r;~ \dot p_r=-\pdv{\widetilde H}{r};~ \dot\theta=\frac{p_\theta}{r^2};~\dot p_\theta=-\pdv{\widetilde H}{\theta};\\
    &\dot\phi=\frac{p_\phi}{r^2\sin^2\theta}-\mu;\\
    &\dot p_\phi=-\pdv{\widetilde H}{\phi}+\mu r^2\gamma\sin^2\theta-\gamma p_\phi+r\sin\theta\sqrt{\gamma T}\xi(t). \label{eq:sm_3D_noncons}
\end{aligned}
\end{equation}

\section{\label{app:unc_noncentral}Uncertainties in \boldmath\texorpdfstring{$\phi$}{phi} - dependent Potentials}
In \cite{supp}, we show that 
\begin{equation}
\begin{aligned}
    &\ev{p_x}=-\mu\ev{y},~~ \ev{p_y}=\mu\ev{x},\\
    &\ev{p_x^2}=T+\mu^2\ev{y^2},~~ \ev{p_y^2}=T+\mu^2\ev{x^2}.
\end{aligned}
\end{equation}
We further show in \cite{supp} that $\mu=\ev{\frac{p_\phi}{r^2}}=\frac{\ev{p_\phi}}{\ev{r^2}}$ using which
\begin{equation}
        \Delta p_x^2=T+\frac{\ev{p_\phi}^2}{\ev{r^2}^2}\Delta y^2~~\mathrm{and}~~
    \Delta p_y^2=T+\frac{\ev{p_\phi}^2}{\ev{r^2}^2}\Delta x^2.
\end{equation}
Thus, 
\begin{equation}
\begin{aligned}
    &(\Delta x\Delta p_x)^2=T\Delta x^2+\ev{p_\phi}^2\qty(\frac{\Delta x\Delta y}{\ev{r^2}})^2,\\
    &(\Delta y\Delta p_y)^2=T\Delta y^2+\ev{p_\phi}^2\qty(\frac{\Delta x\Delta y}{\ev{r^2}})^2.
\end{aligned}
\end{equation}
Therefore,
\begin{equation}
\begin{aligned}
    &\Delta x\Delta p_x\geq \min\qty(\frac{\Delta x\Delta y}{\ev{r^2}})\abs{\ev{p_\phi}}
    ~~\mathrm{and}\\
    &\Delta y\Delta p_y\geq \min\qty(\frac{\Delta x\Delta y}{\ev{r^2}})\abs{\ev{p_\phi}}.
\end{aligned}
\end{equation}
The equalities are respected when
\begin{equation}
    \min\qty(\frac{\Delta x\Delta y}{\ev{r^2}})=\qty(\frac{\Delta x\Delta y}{\ev{r^2}})(T\to0).
\end{equation}
We define $c_{xx}=c_{yy}=c=\min\qty(\frac{\Delta x\Delta y}{\ev{r^2}})$. We now find a bound on $c$. We first use the simple relation
\begin{equation}
    (\Delta x-\Delta y)^2\geq0\imp \Delta x^2+\Delta y^2-2\Delta x\Delta y\geq0.
\end{equation}
Using this,
\begin{equation}
   \frac{\Delta x\Delta y}{\ev{r^2}}\leq\frac{\Delta x^2+\Delta y^2}{2\ev{r^2}}=\frac{\ev{r^2}-\ev{r\cos\phi}^2-\ev{r\sin\phi}^2}{2\ev{r^2}}.
\end{equation}
Now, using the Cauchy-Schwarz inequality, we have $\ev{r\cos\phi}^2\leq\ev{r^2}\ev{\cos^2\phi}$ and $\ev{r\sin\phi}^2\leq\ev{r^2}\ev{\sin^2\phi}$. We therefore end up with
\begin{equation}
    \frac{\ev{r\cos\phi}^2+\ev{r\sin\phi}^2}{\ev{r^2}}=\frac{\ev{x}^2+\ev{y}^2}{\ev{r^2}}\leq1.
\end{equation}
Using this,
\begin{equation}
    \frac{\Delta x\Delta y}{\ev{r^2}}\leq\frac{1}{2}\qty(1-\frac{\ev{r\cos\phi}^2+\ev{r\sin\phi}^2}{\ev{r^2}})\leq\frac{1}{2}.
\end{equation}
The above implies
\begin{equation}
    c=\min\qty(\frac{\Delta x\Delta y}{\ev{r^2}})\leq\frac{1}{2}.
\end{equation}
The other relations we have are
\begin{equation}
\begin{aligned}
    &(\Delta x\Delta p_y)^2=T\Delta x^2+\ev{p_\phi}^2\qty(\frac{\Delta x^2}{\ev{r^2}})^2,\\
    &(\Delta y\Delta p_x)^2=T\Delta y^2+\ev{p_\phi}^2\qty(\frac{\Delta y^2}{\ev{r^2}})^2. 
\end{aligned}
\end{equation}
Thus,
\begin{equation}
\begin{aligned}
    &\Delta x\Delta p_y\geq \min\qty(\frac{\Delta x^2}{\ev{r^2}})\abs{\ev{p_\phi}}~~
    \mathrm{and}\\
    &\Delta y\Delta p_x\geq \min\qty(\frac{\Delta y^2}{\ev{r^2}})\abs{\ev{p_\phi}}.
\end{aligned}
\end{equation}
The equalities hold when
\begin{equation}
\begin{aligned}
    &\min\qty(\frac{\Delta x^2}{\ev{r^2}})=\qty(\frac{\Delta x^2}{\ev{r^2}})(T\to0)~~\mathrm{and}\\
    &\min\qty(\frac{\Delta y^2}{\ev{r^2}})=\qty(\frac{\Delta y^2}{\ev{r^2}})(T\to0).
\end{aligned}
\end{equation}
Define $c_{xy}=\min\qty(\frac{\Delta x^2}{\ev{r^2}})$ and $c_{yx}=\min\qty(\frac{\Delta y^2}{\ev{r^2}})$. We find a bound on $c_{xy}$ and $c_{yx}$. Suppose $\Delta x^2>\ev{r^2}$. Then we have $\ev{r^2\cos^2\phi}>\ev{r^2}+\ev{r\cos\phi}^2=\ev{r^2\cos^2\phi}+\ev{r^2\sin^2\phi}+\ev{r\cos\phi}^2$. This leads to $\ev{r^2\sin^2\phi}+\ev{r\cos\phi}^2<0$, which is clearly incorrect. Thus, $\Delta x^2\leq\ev{r^2}$. A similar argument shows $\Delta y^2\leq \ev{r^2}$. Therefore, we have $c_{xy},c_{yx}\in[0,1]$.

\bibliography{references.bib}

\end{document}


\title{Supplemental Material for ``Position-Momenta Uncertainties  in   Classical Systems"}
\author{Dipesh K. Singh} \email{dks20ms176@iiserkol.ac.in} \author{P. K. Mohanty}
\email{pkmohanty@iiserkol.ac.in}
\affiliation {Department of Physical Sciences, Indian Institute of Science Education and Research Kolkata, Mohanpur, 741246, India.}
\date{\today}

\begin{abstract}
In this Supplemental Material, we explicitly show the relationship between steady state moments used in the main text. We also justify the low temperature behaviour of the radial moments as well as the energy and angular momentum fluctuations. We present full calculations of steady state features of 2D SHO in the non-conserved ensemble. We finally present some results in 3D.
\end{abstract}
\maketitle

\section{\label{sec:sm_momentrelation}Relationship between Moments}
Consider the general $\phi$-dependent potential. The partition function is obtained as follows.
\begin{align*}
    \Z(\beta,\mu)&=\int_{-\infty}^{\infty}\dd{p_r} \exp(-\frac{\beta p_r^2}{2})\int_{0}^{2\pi}\dd{\phi}\int_{0}^{\infty}\dd{r}\int_{-\infty}^{\infty}\dd{p_\phi} \exp(-\frac{\beta}{2r^2}(p_\phi^2-2r^2\mu p_\phi))\exp(-\beta V(r,\phi))\\
    &=\sqrt{\frac{2\pi}{\beta}}\int_{0}^{2\pi}\dd{\phi}\int_{0}^{\infty}\dd{r} \int_{-\infty}^{\infty}\dd{p_\phi}\exp(-\frac{\beta}{2r^2}(p_\phi-\mu r^2)^2)\exp(-\beta \qty(V(r,\phi)-\frac{\mu^2r^2}{2}))\\
    \imp \Z(\beta,\mu)&=\frac{2\pi}{\beta}\int_{0}^{2\pi}\dd{\phi}\int_{0}^{\infty}\dd{r} r\exp(-\beta \qty(V(r,\phi)-\frac{\mu^2r^2}{2})). \numberthis
\end{align*}
We can therefore write
\begin{equation}
    \ev{r^n}=\dfrac{\displaystyle\int_{0}^{2\pi}\dd{\phi}\int_{0}^{\infty}\dd{r} r^{n+1}\exp(-\beta \qty(V(r,\phi)-\frac{\mu^2r^2}{2}))}{\displaystyle\int_{0}^{2\pi}\dd{\phi}\int_{0}^{\infty}\dd{r} r\exp(-\beta \qty(V(r,\phi)-\frac{\mu^2r^2}{2}))}.
\end{equation}
From above, it is clear that 
\begin{equation}
    \ev{p_\phi}=\frac{1}{\Z}\pdv{\Z}{(\beta\mu)}=\frac{\displaystyle\int_{0}^{2\pi}\dd{\phi}\int_{0}^{\infty}\dd{r} \mu r^{3}\exp(-\beta \qty(V(r,\phi)-\frac{\mu^2r^2}{2}))}{\displaystyle\int_{0}^{2\pi}\dd{\phi}\int_{0}^{\infty}\dd{r} r\exp(-\beta \qty(V(r,\phi)-\frac{\mu^2r^2}{2}))}=\mu\ev{r^2}.
\end{equation}
Moreover,
\begin{align*}
    \ev{\frac{p_\phi}{r^2}}&=\frac{\displaystyle\int_{0}^{2\pi}\dd{\phi}\int_{0}^{\infty}\dd{r} \int_{-\infty}^{\infty}\dd{p_\phi}\frac{p_\phi}{r^2}\exp(-\frac{\beta}{2r^2}(p_\phi-\mu r^2)^2)\exp(-\beta \qty(V(r,\phi)-\frac{\mu^2r^2}{2}))}{\displaystyle\int_{0}^{2\pi}\dd{\phi}\int_{0}^{\infty}\dd{r} \int_{-\infty}^{\infty}\dd{p_\phi}\exp(-\frac{\beta}{2r^2}(p_\phi-\mu r^2)^2)\exp(-\beta \qty(V(r,\phi)-\frac{\mu^2r^2}{2}))}\\
    &=\frac{\displaystyle\int_{0}^{2\pi}\dd{\phi}\int_{0}^{\infty}\dd{r} \int_{-\infty}^{\infty}\dd{p_\phi}\frac{p_\phi+\mu r^2}{r^2}\exp(-\frac{\beta p_\phi^2}{2r^2})\exp(-\beta \qty(V(r,\phi)-\frac{\mu^2r^2}{2}))}{\displaystyle\int_{0}^{2\pi}\dd{\phi}\int_{0}^{\infty}\dd{r} \int_{-\infty}^{\infty}\dd{p_\phi}\exp(-\frac{\beta p_\phi^2}{2r^2})\exp(-\beta \qty(V(r,\phi)-\frac{\mu^2r^2}{2}))}\\
    \imp \ev{\frac{p_\phi}{r^2}}&=\mu. \numberthis
\end{align*}
Therefore in steady state,
\begin{equation}
    \mu=\frac{\ev{p_\phi}}{\ev{r^2}}=\ev{\frac{p_\phi}{r^2}}.
\end{equation}
Thus, existence of $\ev{r^2}$ in steady state is essential for the existence of $\ev{p_\phi}$ and hence, for the possibility of tuning $p_\phi$. We also compute $\ev{p_\phi^2}$ to be used in the next section.
\begin{equation}
    \ev{p_\phi^2}=\frac{1}{\beta^2\Z}\pdv[2]{\Z}{\mu}=T\ev{r^2}+\mu^2\ev{r^4}.
\end{equation}

The $p_{x,y}$ can be expressed as follows
\begin{equation}
    p_x=p_r\cos\phi-\frac{p_\phi}{r}\sin\phi \qand p_y=p_r\sin\phi+\frac{p_\phi}{r}\cos\phi
\end{equation}
It is a trivial exercise to show $\ev{p_r}=0=\ev{\frac{p_rp_\phi}{r}}$ and $\ev{p_r^2}=T$. Thus,
\begin{equation}
    \ev{p_x}=-\ev{\frac{p_\phi}{r}\sin\phi},~ \ev{p_y}=\ev{\frac{p_\phi}{r}\cos\phi},~ \ev{p_x^2}=T\ev{\cos^2\phi}+\ev{\frac{p_\phi^2}{r^2}\sin^2\phi},~ \ev{p_y^2}=T\ev{\sin^2\phi}+\ev{\frac{p_\phi^2}{r^2}\cos^2\phi}.
\end{equation}
We compute $\ev{\frac{p_\phi}{r}g(\phi)}$ and $\ev{\frac{p_\phi^2}{r^2}g(\phi)}$ for some function $g(\phi)$.
\begin{align*}
    \ev{\frac{p_\phi}{r}g(\phi)}&=\frac{\displaystyle\int_{0}^{2\pi}\dd{\phi}\int_{0}^{\infty}\dd{r} \int_{-\infty}^{\infty}\dd{p_\phi}\frac{p_\phi}{r}g(\phi)\exp(-\frac{\beta}{2r^2}(p_\phi-\mu r^2)^2)\exp(-\beta \qty(V(r,\phi)-\frac{\mu^2r^2}{2}))}{\displaystyle\int_{0}^{2\pi}\dd{\phi}\int_{0}^{\infty}\dd{r} \int_{-\infty}^{\infty}\dd{p_\phi}\exp(-\frac{\beta}{2r^2}(p_\phi-\mu r^2)^2)\exp(-\beta \qty(V(r,\phi)-\frac{\mu^2r^2}{2}))}\\
    &=\frac{\displaystyle\int_{0}^{2\pi}\dd{\phi}\int_{0}^{\infty}\dd{r} \int_{-\infty}^{\infty}\dd{p_\phi}\qty(\frac{p_\phi+\mu r^2}{r}) g(\phi)\exp(-\frac{\beta p_\phi^2}{2r^2})\exp(-\beta \qty(V(r,\phi)-\frac{\mu^2r^2}{2}))}{\displaystyle\int_{0}^{2\pi}\dd{\phi}\int_{0}^{\infty}\dd{r} \int_{-\infty}^{\infty}\dd{p_\phi}\exp(-\frac{\beta p_\phi^2}{2r^2})\exp(-\beta \qty(V(r,\phi)-\frac{\mu^2r^2}{2}))}\\
    \imp \ev{\frac{p_\phi}{r}g(\phi)}&=\mu\ev{r g(\phi)}. \numberthis
\end{align*}
Using this,
\begin{equation}
    \ev{p_x}=-\mu\ev{r\sin\phi}=-\mu\ev{y}\qand \ev{p_y}=\mu\ev{r\cos\phi}=\mu\ev{x}.
\end{equation}
Similarly,
\begin{align*}
    \ev{\frac{p_\phi^2}{r^2}g(\phi)}&=\frac{\displaystyle\int_{0}^{2\pi}\dd{\phi}\int_{0}^{\infty}\dd{r} \int_{-\infty}^{\infty}\dd{p_\phi}\frac{p_\phi^2}{r^2}g(\phi)\exp(-\frac{\beta}{2r^2}(p_\phi-\mu r^2)^2)\exp(-\beta \qty(V(r,\phi)-\frac{\mu^2r^2}{2}))}{\displaystyle\int_{0}^{2\pi}\dd{\phi}\int_{0}^{\infty}\dd{r} \int_{-\infty}^{\infty}\dd{p_\phi}\exp(-\frac{\beta}{2r^2}(p_\phi-\mu r^2)^2)\exp(-\beta \qty(V(r,\phi)-\frac{\mu^2r^2}{2}))}\\
    &=\frac{\displaystyle\int_{0}^{2\pi}\dd{\phi}\int_{0}^{\infty}\dd{r} \int_{-\infty}^{\infty}\dd{p_\phi}\frac{(p_\phi+\mu r^2)^2}{r^2} g(\phi)\exp(-\frac{\beta p_\phi^2}{2r^2})\exp(-\beta \qty(V(r,\phi)-\frac{\mu^2r^2}{2}))}{\displaystyle\int_{0}^{2\pi}\dd{\phi}\int_{0}^{\infty}\dd{r} \int_{-\infty}^{\infty}\dd{p_\phi}\exp(-\frac{\beta p_\phi^2}{2r^2})\exp(-\beta \qty(V(r,\phi)-\frac{\mu^2r^2}{2}))}\\
    &=\frac{\displaystyle\int_{0}^{2\pi}\dd{\phi}\int_{0}^{\infty}\dd{r} \int_{-\infty}^{\infty}\dd{p_\phi}\qty(\frac{p_\phi^2}{r^2}+\mu^2r^2)g(\phi)\exp(-\frac{\beta p_\phi^2}{2r^2})\exp(-\beta \qty(V(r,\phi)-\frac{\mu^2r^2}{2}))}{\displaystyle\int_{0}^{2\pi}\dd{\phi}\int_{0}^{\infty}\dd{r} \int_{-\infty}^{\infty}\dd{p_\phi}\exp(-\frac{\beta p_\phi^2}{2r^2})\exp(-\beta \qty(V(r,\phi)-\frac{\mu^2r^2}{2}))}\\
    \imp \ev{\frac{p_\phi^2}{r^2}g(\phi)}&=T\ev{g(\phi)}+\mu^2\ev{r^2 g(\phi)}. \numberthis
\end{align*}
Using this,
\begin{equation}
\begin{aligned}
    \ev{p_x^2}&=T\ev{\cos^2\phi}+T\ev{\sin^2\phi}+\mu^2\ev{r^2\sin^2\phi}=T+\mu^2\ev{y^2}\\
    \ev{p_y^2}&=T\ev{\sin^2\phi}+T\ev{\cos^2\phi}+\mu^2\ev{r^2\cos^2\phi}=T+\mu^2\ev{x^2}
\end{aligned}
\end{equation}
Thus,
\begin{equation}
\begin{aligned}
    \Delta p_x^2&=T+\mu^2\ev{y^2}-\mu^2\ev{y}^2=T+\mu^2\Delta y^2=T+\frac{\ev{p_\phi}^2}{\ev{r^2}^2}\Delta y^2\\
    \Delta p_y^2&=T+\mu^2\ev{x^2}-\mu^2\ev{x}^2=T+\mu^2\Delta x^2=T+\frac{\ev{p_\phi}^2}{\ev{r^2}^2}\Delta x^2
\end{aligned}
\end{equation}
For central potentials, we have $\ev{x}=\ev{y}=0$ and $\ev{x^2}=\ev{y^2}=\frac{1}{2}\ev{r^2}$. So the relations above simplify to
\begin{equation}
    \ev{p_x}=\ev{p_y}=0 \qand \ev{p_x^2}=\ev{p_y^2}=T+\frac{\ev{p_\phi}^2}{2\ev{r^2}}.
\end{equation}

\section{\label{sec:sm_saddlepoint}Calculation of Radial Moments for Steady States of Central Potentials}
\subsection{Conserved Angular Momentum}
For the distribution $e^{-\beta H}$ with $H=p_r^2/2+\ell^2/2r^2+V(r)$, the radial moments for $n\geq0$ are given by
\begin{equation}
    \ev{r^n}=\dfrac{\displaystyle\int_0^\infty\dd{r} r^n\ \exp(-\beta \widetilde{V}(r))}{\displaystyle\int_0^\infty\dd{r} \exp(-\beta \widetilde{V}(r))};\quad \widetilde{V}(r)=\dfrac{\ell^2}{2r^2}+V(r).
\end{equation}
$\widetilde{V}(r)$ is the effective radial potential. In the $T\to0$ limit, (equivalently, $\beta\to\infty$), the value of the integral is dominated by the strongest minimum. Let $r=r_c$ denote this minimum. Performing the saddle-point approximation around $r=r_c$, the integral becomes
\begin{equation}
    \ev{r^n}\sim\dfrac{\displaystyle\int_0^\infty\dd{r} r^n\ \exp(- \frac{\beta\widetilde{V}''(r_c)}{2}(r-r_c)^2)}{\displaystyle\int_0^\infty\dd{r} \exp(- \frac{\beta\widetilde{V}''(r_c)}{2}(r-r_c)^2)}=\dfrac{\displaystyle\int_{-r_c}^\infty\dd{r} (r+r_c)^n\ \exp(- \frac{\beta\widetilde{V}''(r_c)r^2}{2})}{\displaystyle\int_{-r_c}^\infty\dd{r} \exp(- \frac{\beta\widetilde{V}''(r_c)r^2}{2})}.
\end{equation}
The above Gaussian can be integrated in terms of confluent hypergeometric function ${}_1F_{1}(a;b;z)$ and Gamma function. The integral evaluates to
\begin{equation}
    \ev{r^n}\sim\frac{b^{-\frac{n}{2}} \left(\Gamma\qty(\frac{n+1}{2}) \, _1F_1\left(-\frac{n}{2};\frac{1}{2};-r_c^2 b\right)+r_c n\sqrt{b}\Gamma\qty(\frac{n}{2}) \, _1F_1\left(\frac{1-n}{2};\frac{3}{2};-r_c^2 b\right)\right)}{\sqrt{\pi } \left(\text{erf}\left(r_c \sqrt{b}\right)+1\right)};\ b=\dfrac{\beta \widetilde{V}''(r_c)}{2}.
\end{equation}
The asymptotic form of ${}_1F_1(p;q;z)$ as $z\to-\infty$, when $\Gamma(q-p)$ is finite is given by
\begin{equation}
    {}_1F_1(p;q;z)\sim \dfrac{\Gamma(q)}{\Gamma(q-p)}(-z)^{-p}.
\end{equation}
Finally taking the limit $\beta\to\infty$ (equivalently, $b\to\infty$),
\begin{equation}
    \ev{r^n}\sim \dfrac{b^{-n/2}}{2\sqrt{\pi}}\qty(\dfrac{\Gamma\qty(\frac{n+1}{2})\Gamma\qty(\frac{1}{2})r_c^n\ b^{n/2}}{\Gamma\qty(\frac{n+1}{2})}+\dfrac{n\Gamma\qty(\frac{n}{2})\Gamma(\frac{3}{2})r_c^nb^{n/2}}{\Gamma\qty(\frac{n+2}{2})})
\end{equation}
which simplifies to 
\begin{equation}
    \ev{r^n}(\beta\to\infty)=r_c^n.
\end{equation}
Thus, the particle is definitely on $r=r_c$ as $T\to0$ and $\Delta r(T\to0)=0$ and $\Delta E(T\to0)=0$.

\subsection{Non-Conserved Angular Momentum}
In the non-conserved ensemble, the radial moments for $n\geq0$ are given by
\begin{equation}
    \ev{r^n}=\dfrac{\displaystyle\int_0^\infty\dd{r} r^{n+1}\ \exp(-\beta f(r))}{\displaystyle\int_0^\infty\dd{r} r \exp(-\beta f(r))};\quad f(r)=V(r)-\dfrac{\mu^2r^2}{2}
\end{equation}
Suppose $f(r)$ has the strongest minimum at $r=r_c$. Performing the saddle-point approximation about $r=r_c$, we have
\begin{equation}
    \ev{r^n}\sim\dfrac{\displaystyle\int_0^\infty\dd{r} r^{n+1}\ \exp(- \frac{\beta f''(r_c)}{2}(r-r_c)^2)}{\displaystyle\int_0^\infty\dd{r} r\exp(- \frac{\beta f''(r_c)}{2}(r-r_c)^2)}=\dfrac{\displaystyle\int_{-r_c}^\infty\dd{r} (r+r_c)^{n+1}\ \exp(- \frac{\beta f''(r_c)r^2}{2})}{\displaystyle\int_{-r_c}^\infty\dd{r} (r+r_c)\exp(- \frac{\beta f''(r_c)r^2}{2})} \label{radialnoncons}
\end{equation}
Proceeding as before, we obtain
\begin{equation}
    \ev{r^n}(\beta\to\infty)=\dfrac{r_c^{n+1}}{r_c}=r_c^n \label{saddlenoncons}
\end{equation}
Consider the potential $V(r)=kr^\alpha/\alpha$, $\alpha\geq2$. Thus, $f(r)=kr^\alpha/\alpha-\mu^2r^2/2$. The condition of minimum reduces to 
\begin{align}
    f'(r_c)=0 \imp r_c\qty(kr_c^{\alpha-2}-\mu^2)=0\qcomma f''(r_c)>0 \imp k(\alpha-1)r_c^{\alpha-2}-\mu^2>0
\end{align}
Consider the marginal case $\alpha=2$. Convergence requires $\mu^2<k$ and $r_c=0$ is the minimum of $f(r)$. Eq. \eqref{radialnoncons} becomes exact, however, the result in Eq. \eqref{saddlenoncons} becomes inaccurate. On computing the Gaussian integrals (see next section for full calculation), one finds that for $\ev{p_\phi}\neq0$, $\mu^2\to k$ as $T\to0$, i.e., $f''(r_c)\to0$ as $\beta\to\infty$. Thus, the results from the saddle-point approximation are not applicable in this case. $\alpha>2$ potentials are exempt from bounds on $\mu$ for convergence. There is a minimum $r_c\neq0$ and $f''(r_c)$ remains positive as $\beta\to0$. We are free to use result \eqref{saddlenoncons} here. In fact, using the relation, $\ev{p_\phi}=\mu\ev{r^2}=\mu r_c^2$ as $T\to0$, we find $r_c$ to be $r_c=\qty(\frac{\ev{p_\phi}^2}{k})^{\frac{1}{\alpha+2}}$ and $f''(r_c)=\frac{\ev{p_\phi}^2}{r_c^4}(\alpha-2)$ which is indeed positive for $\alpha>2$. Moreover, the particle is definitely on $r=r_c$ as $T\to0$ and $\Delta r(T\to0)=0=\Delta E(T\to0)$. Also, $\Delta p_\phi^2(T\to0)=(\mu^2\ev{r^4}-\mu^2\ev{r^2}^2)(T\to0)=0$.

\section{Steady State Properties of 2D SHO in the Non-Conserved Ensemble}
The partition function reads 
\begin{equation}
    \mathcal{Z}(\beta,\mu)=\frac{4\pi^2}{\beta}\int_{0}^{\infty}\dd{r}\exp(-\frac{\beta}{2}\qty(k-\mu^2)r^2)=\dfrac{4\pi^2}{\beta^2(k-\mu^2)}
\end{equation}
where we require
\begin{equation}
    \mu^2<k\imp -\sqrt{k}<\mu<\sqrt{k}
\end{equation}
for convergence. We have the following moments,
\begin{align}
    \ev{E^{n}}&=\dfrac{T^n\ n!(-1)^n}{2\mu}\qty(\qty(\mu+\sqrt{k})\qty(\dfrac{\sqrt{k}}{\mu-\sqrt{k}})^n+\qty(\mu-\sqrt{k})\qty(\dfrac{-\sqrt{k}}{\mu+\sqrt{k}})^n) \label{Eavg}\\
    \ev{r^n}&=\Gamma\qty(\dfrac{n}{2}+1)\qty(\dfrac{2T}{k-\mu^2})^{\frac{n}{2}}\\
    \ev{p_\phi^n}&=\dfrac{n!\ T^n}{2\sqrt{k}}\dfrac{\qty(\mu+\sqrt{k})^{n+1}-\qty(\mu-\sqrt{k})^{n+1}}{\qty(k-\mu^2)^{n}}\\
    \ev{p_r^n}&=\begin{cases}
        0, &n=1,3,5,7,\ldots\\
        T^{\frac{n}{2}}(n-1)!!, &n=2,4,6,8,\ldots
    \end{cases}
\end{align}
In particular, for energy, 
\begin{equation}
    \ev{\dfrac{p_r^2}{2}}=\dfrac{T}{2}\qcomma \ev{\dfrac{p_\phi^2}{2r^2}}=\dfrac{T}{2}\dfrac{k+\mu^2}{k-\mu^2}\qand \ev{\dfrac{1}{2}kr^2}=\dfrac{kT}{k-\mu^2} \imp \ev{E}=\dfrac{2kT}{k-\mu^2}
\end{equation}
We also have, for $E$,
\begin{equation}
    \ev{E}=\dfrac{2kT}{k-\mu^2}\qcomma\ev{E^2}=\dfrac{2kT^2\qty(\mu^2+3k)}{(k-\mu^2)^2}\imp \ev{E^2}-\ev{E}^2=\dfrac{2kT^2\qty(k+\mu^2)}{\qty(k-\mu^2)^2},
\end{equation}
for $r$,
\begin{equation}
    \ev{r}=\sqrt{\dfrac{\pi T}{2(k-\mu^2)}}\qcomma\ev{r^2}=\dfrac{2T}{k-\mu^2}=\frac{4}{\pi}\ev{r}^2\imp \ev{r^2}-\ev{r}^2=\dfrac{T}{2(k-\mu^2)}(4-\pi)=\qty(\frac{4}{\pi}-1)\ev{r}^2,
\end{equation}
and for $p_\phi$,
\begin{equation}
    \ev{p_\phi}=\dfrac{2\mu T}{k-\mu^2}\qcomma \ev{p_\phi^2}=\dfrac{2(3\mu^2+k)}{(k-\mu^2)^2}\ T^2 \imp \ev{p_\phi^2}-\ev{p_\phi}^2=\dfrac{2(k+\mu^2)}{(k-\mu^2)^2}\ T^2.
\end{equation}
We can always find $\mu\in\qty(-\sqrt{k},\sqrt{k})$ such that $\ev{p_\phi}=\ell$. The $\mu$ is given by
\begin{equation} 
    \mu=\dfrac{-T+\sqrt{T^2+k\ell^2}}{\ell} \qand \ell=0~\Leftrightarrow~\mu=0.
\end{equation}
Using the above $\mu$, we have
\begin{equation}
    \ev{\dfrac{p_r^2}{2}}=\dfrac{T}{2}\qcomma \ev{\dfrac{p_\phi^2}{2r^2}}=\dfrac{1}{2}\sqrt{T^2+k\ell^2}\qand \ev{\dfrac{1}{2}kr^2}=\dfrac{T}{2}+\dfrac{1}{2}\sqrt{T^2+k\ell^2}
\end{equation}
with
\begin{equation}
    \ev{E}=T+\sqrt{T^2+k\ell^2}\qcomma \ev{E^2}=3T^2+3T\sqrt{k l_0^2+T^2}+2k\ell^2
\end{equation}
so
\begin{equation}
    \ev{E^2}-\ev{E}^2=T^2+T\sqrt{k l_0^2+T^2}+k \ell^2.
\end{equation}
\begin{table}[ht]
    \centering
    \begin{tabularx}{\textwidth}{|>{\centering\arraybackslash}X|>{\centering\arraybackslash}X|}
        \hline
        $\alpha=2$ & $\alpha>2$\\ \hline
        &\\
        $\mu\in\qty(-\sqrt{k},\sqrt{k})$& $\mu\in(-\infty,\infty)$ \\ &\\ \hline &\\
        $\mu(T\to0)=\text{sgn}(\ev{p_\phi})\sqrt{k}$ & $\mu(T\to0)=\dfrac{\ev{p_\phi}}{r_c^2}$ \\ &\\ \hline 
        &\\
        $\ev{r}(T\to0)=\frac{\sqrt{\pi}}{2}\ r_c$& $\ev{r}(T\to0)=r_c$ \\ &\\ \hline &\\
        $\Delta r(T\to0)=r_c\sqrt{\frac{4}{\pi}-1}$ & $\Delta r(T\to0)=0$ \\ &\\ \hline &\\ 
        $\Delta{E}(T\to0)=\sqrt{k}\abs{\ev{p_\phi}}\qcomma C_V=\frac{\Delta E^2}{T^2}\to\infty$ as $T\to0$ & $\Delta{E}(T\to0)=0$\\ & \\ \hline&\\
        $\Delta p_\phi(T\to0)=\abs{\ev{p_\phi}}$ & $\Delta p_\phi(T\to0)=0$ \\&\\\hline
    \end{tabularx}
        \caption{Difference between $\alpha=2$ and $\alpha>2$ potentials. $r_c=\qty(\frac{\ev{p_\phi}^2}{k})^{\frac{1}{\alpha+2}}$.}
        \label{table:diff_alpha}
\end{table}
Thus, the specific heat,
\begin{equation}
    C_V=\dfrac{\ev{E^2}-\ev{E}^2}{T^2}=1+\dfrac{\sqrt{k\ell^2+T^2}}{T}+\dfrac{k\ell^2}{T^2} \to\infty \qas T\to0.
\end{equation}
For $r$, we now have
\begin{equation}
    \ev{r}=\dfrac{1}{2}\sqrt{\dfrac{\pi}{k}}\sqrt{T+\sqrt{T^2+k\ell^2}}\qcomma \ev{r^2}=\dfrac{T+\sqrt{T^2+k\ell^2}}{k} \imp \ev{r^2}-\ev{r}^2=\qty(1-\dfrac{\pi}{4})\dfrac{T+\sqrt{T^2+k\ell^2}}{k}.
\end{equation}
Also
\begin{equation}
    \ev{p_\phi^2}=\dfrac{T^2+T\sqrt{k \ell^2+T^2}}{k}+2 \ell^2 \imp \ev{p_\phi^2}-\ev{p_\phi}^2=\ell^2+\dfrac{T \left(\sqrt{k \ell^2+T^2}+T\right)}{k}.
\end{equation}
The variances do not vanish in the limit $T\to0$. 

For $N$ such oscillators, we have 
\begin{equation}
    \ev{p_\phi^2}-\ev{p_\phi}^2=\dfrac{1}{\beta^2}\pdv[2]{\mu}(\ln\mathcal{Z}^N)=2NT^2\ \dfrac{k+\mu^2}{\qty(k-\mu^2)^2} \qand \ev{p_\phi}=\dfrac{1}{\beta}\pdv{\mu}(\ln\mathcal{Z}^N)=\dfrac{2N\mu T}{k-\mu^2}.
\end{equation}
Thus,
\begin{equation}
    \dfrac{\ev{p_\phi^2}-\ev{p_\phi}^2}{\ev{p_\phi}^2}=\dfrac{1}{N}\dfrac{k+\mu^2}{2\mu^2}\to0 \qq{as} N\to\infty.
\end{equation}
Some other relations between the variances of $E$, $r$ and $p_\phi$ are
\begin{equation}
    \Delta E^2=k\Delta{p_{\phi}}^2\qand
    \Delta{p_{\phi}}^2=\ell^2+\dfrac{T}{1-\pi/4}\Delta r^2 \imp 
    \Delta {E}^2=k\ell^2+\dfrac{kT}{1-\pi/4}\Delta r^2.
\end{equation}
We summarize the differences between $\alpha=2$ and $\alpha>2$ potentials in Table \ref{table:diff_alpha}.

\section{3D Central Potentials} 
\subsection{Conserved Angular Momentum}
We consider the Hamiltonian of particle in a 3D central potential in spherical polar coordinates,
\begin{equation}
    H=\frac{p_r^2}{2}+\frac{p_\theta^2}{2r^2}+\frac{p_\phi^2}{2r^2\sin^2\theta}+V(r).
\end{equation}
In spherical polar coordinates $(r,\theta,\phi)$,
\begin{equation}
    x=r\sin\theta\cos\phi\qcomma y=r\sin\theta\sin\phi\qand z=r\cos\theta.
\end{equation}
Using $\dot x=p_x$, $\dot y=p_y$ and $\dot z=p_z$, we get the following relation between $\qty{p_x,p_y,p_z}$ and $\qty{p_r,p_\theta,p_\phi}$
\begin{equation}
\begin{aligned}
    p_x&=p_r\sin\theta\cos\phi+\frac{p_\theta}{r}\cos\theta\cos\phi-\frac{p_\phi}{r\sin\theta}\sin\phi\\
    p_y&=p_r\sin\theta\sin\phi+\frac{p_\theta}{r}\cos\theta\sin\phi+\frac{p_\phi}{r\sin\theta}\cos\phi\\
    p_z&=p_r\cos\theta-\frac{p_\theta}{r}\sin\theta
\end{aligned}
\end{equation}
In the conserved ensemble, $p_\phi=\ell=\mathrm{constant}$. We have the following equations of motion
\begin{equation}
    \dot r=p_r;~ \dot p_r=-\pdv{V}{r}+\frac{p_\theta^2}{r^3}+\frac{\ell^2}{r^3\sin^2\theta}-\gamma p_r+\sqrt{\gamma T}\xi(t);~ \dot\theta=\frac{p_\theta}{r^2};~ \dot p_\theta=\frac{\ell^2}{r^2}\cot\theta\csc^2\theta;~\dot\phi=\frac{\ell^2}{r^2\sin^2\theta}
\end{equation}
with $\ev{\xi(t)}=0$ and $\ev{\xi(t)\xi(t')}=2\delta(t-t')$. The Fokker-Planck equation reads
\begin{equation}
    \pdv{P}{t}=[H,P]+\pdv{p_r}(\gamma p_rP+\gamma T\pdv{P}{p_r})
\end{equation}
with the steady state solution, $P(r,p_r,\theta,p_\theta,\ell)=\frac{\exp(-\beta H)}{\Z_\ell(\beta)}$. Here
\begin{equation}
    \Z_\ell(\beta)=\int_{0}^{2\pi}\dd{\phi}\int_{0}^{\infty}\dd{r}\int_{0}^{\pi}\dd{\theta}\int_{-\infty}^{\infty}\dd{p_r}\int_{-\infty}^{\infty}\dd{p_\theta}\exp(-\beta\qty(\frac{p_r^2}{2}+\frac{p_\theta^2}{2r^2}+\frac{\ell^2}{2r^2\sin^2\theta}+V(r)))
\end{equation}
needs to be finite for $P$ to be a valid PDF. On integrating, the partition function can be written as 
\begin{equation}
    \Z_\ell(\beta)=\frac{4\pi^2}{\beta}\int_{0}^{\infty}\dd{r}r\int_{0}^{\pi}\dd{\theta}\exp(-\frac{\beta\ell^2}{2r^2\sin^2\theta})\exp(-\beta V(r))=\frac{4\pi^3}{\beta}\int_{0}^{\infty}\dd{r}r\ \erfc{\sqrt{\frac{\beta\ell^2}{2}}\frac{1}{r}}\exp(-\beta V(r))
\end{equation}
where $\erfc{z}=1-\erf(z)$ is the complementary error function. As the steady state distribution is $\phi$ independent, using $\ev{\sin\phi}=\ev{\cos\phi}=0$ and $\ev{\sin^2\phi}=\ev{\cos^2\phi}=\frac{1}{2}$, 
\begin{equation}
    \ev{x}=\ev{y}=0 \qand \ev{x^2}=\ev{y^2}=\frac{\ev{r^2\sin^2\theta}}{2}.
\end{equation}
$\ev{z^2}=\ev{r^2\cos^2\theta}=\ev{r^2}-\ev{x^2}-\ev{y^2}$ and 
\begin{equation}
    \ev{z}=\frac{\displaystyle\int_{0}^{\infty}\dd{r}r^2\int_{0}^{\pi}\dd{\theta}\cos\theta\exp(-\frac{\beta\ell^2}{2r^2\sin^2\theta})\exp(-\beta V(r))}{\displaystyle\int_{0}^{\infty}\dd{r}r^2\int_{0}^{\pi}\dd{\theta}\exp(-\frac{\beta\ell^2}{2r^2\sin^2\theta})\exp(-\beta V(r))}=0
\end{equation}
as $\cos(\pi-\theta)e^{-\frac{\beta\ell^2}{2r^2\sin^2(\pi-\theta)}}=-\cos\theta\ e^{-\frac{\beta\ell^2}{2r^2\sin^2\theta}}$. It is clear that any average linear in $p_\theta$ and $p_r$ vanishes as the steady state distribution is a Gaussian in these variables. Using this along with $\phi$-independence of the steady state distribution, we can compute the momentum moments as follows.
\begin{equation}
    \ev{p_x}=\ev{p_y}=\ev{p_z}=0.
\end{equation}
It is also very easy to see that $\ev{p_r^2}=T$. Thus, 
\begin{equation}
    \ev{p_z^2}=T\ev{\cos^2\theta}+\ev{\frac{p_\theta^2}{r^2}\sin^2\theta}\qcomma \ev{p_x^2}=\ev{p_y^2}=\frac{T}{2}\ev{\sin^2\theta}+\ev{\frac{p_\theta^2\cos^2\theta}{2r^2}}+\ev{\frac{\ell^2}{2r^2\sin^2\theta}}.
\end{equation}
Consider $\ev{p_z^2}$.
\begin{align*}
    \ev{\frac{p_\theta^2}{r^2}\sin^2\theta}&=\frac{\displaystyle\int_{0}^{\infty}\dd{r}\frac{1}{r^2}\int_{0}^{\pi}\dd{\theta}\sin^2\theta\int_{-\infty}^{\infty}\dd{p_\theta}p_\theta^2\exp(-\frac{\beta p_\theta^2}{2r^2})\exp(-\beta\qty(\frac{\ell^2}{2r^2\sin^2\theta})+V(r))}{\displaystyle\int_{0}^{\infty}\dd{r}\int_{0}^{\pi}\dd{\theta}\int_{-\infty}^{\infty}\dd{p_\theta}\exp(-\frac{\beta p_\theta^2}{2r^2})\exp(-\beta\qty(\frac{\ell^2}{2r^2\sin^2\theta})+V(r))}\\
    &=\frac{1}{\beta}\frac{\displaystyle\int_{0}^{\infty}\dd{r} r\int_{0}^{\pi}\dd{\theta}\sin^2\theta\exp(-\beta\qty(\frac{\ell^2}{2r^2\sin^2\theta})+V(r))}{\displaystyle\int_{0}^{\infty}\dd{r}r\int_{0}^{\pi}\dd{\theta}\exp(-\beta\qty(\frac{\ell^2}{2r^2\sin^2\theta})+V(r))}\\
    \imp \ev{\frac{p_\theta^2}{r^2}\sin^2\theta}&=T\ev{\sin^2\theta}. \numberthis
\end{align*}
Thus, $\ev{p_z^2}=T\ev{\cos^2\theta}+T\ev{\sin^2\theta}=T$. Therefore, $\min(\Delta p_z)=0$. Also, from above, it is clear that $\ev{\frac{p_\theta^2}{2r^2}}=\frac{T}{2}$. Therefore,
\begin{equation}
    \ev{\frac{p_x^2}{2}}+\ev{\frac{p_y^2}{2}}+\ev{\frac{p_z^2}{2}}=\ev{\frac{p_r^2}{2}}+\ev{\frac{p_\theta^2}{2r^2}}+\ev{\frac{\ell^2}{2r^2\sin^2\theta}}\imp \ev{p_x^2}=\ev{p_y^2}=\frac{T}{2}+\ev{\frac{\ell^2}{2r^2\sin^2\theta}}.
\end{equation}
Thus, we have 
\begin{equation}
    \Delta x\Delta p_x=\Delta y\Delta p_y=\frac{1}{2}\sqrt{T\ev{r^2\sin^2\theta}+\ell^2\ev{r^2\sin^2\theta}\ev{\frac{1}{r^2\sin^2\theta}}}.
\end{equation}
If $\ev{r^2}$ exists, $\ev{r^2\sin^2\theta}$ and $\ev{\frac{1}{r^2\sin^2\theta}}$ also exist and the Cauchy–Schwarz inequality ensures $\ev{r^2\sin^2\theta}\ev{\frac{1}{r^2\sin^2\theta}}\geq1$. Thus, we end up with the uncertainty relation
\begin{equation}
    \Delta x\Delta p_x=\Delta y\Delta p_y\geq\frac{\abs{\ell}}{2}.
\end{equation}
The equality holds in the limiting sense as $T\to0$. 

We had found out that $\Delta p_z=\sqrt{T}\to0$ as $T\to0$. We now show that $\Delta z(T\to0)=0$ using the saddle point approximation. We have
\begin{align}
    \ev{z^2}=\ev{r^2\cos^2\theta}&=\frac{\displaystyle\int_{0}^{\infty}\dd{r}r^3\int_{0}^{\pi}\dd{\theta}\cos^2\theta\exp(-\frac{\beta\ell^2}{2r^2\sin^2\theta})\exp(-\beta V(r))}{\displaystyle\int_{0}^{\infty}\dd{r}r\int_{0}^{\pi}\dd{\theta}\exp(-\frac{\beta\ell^2}{2r^2\sin^2\theta})\exp(-\beta V(r))}.
\end{align}
We use
\begin{equation}
\begin{aligned}
    \int_{0}^{\pi}\dd{\theta}\cos^2\theta\exp(-\frac{\beta\ell^2}{2r^2\sin^2\theta})&=\frac{\pi}{2}\qty(1+\frac{\beta\ell^2}{r^2})\erfc{\sqrt{\frac{\beta\ell^2}{2}}\frac{1}{r}}-\sqrt{\frac{\pi\beta\ell^2}{2}}\frac{e^{-\frac{\beta\ell^2}{2r^2}}}{r}\\
    \int_{0}^{\pi}\dd{\theta}\exp(-\frac{\beta\ell^2}{2r^2\sin^2\theta})&=\pi\ \erfc{\sqrt{\frac{\beta\ell^2}{2}}\frac{1}{r}}
\end{aligned}
\end{equation}
Using these, we can write
\begin{equation}
    \ev{z^2}=\frac{\ev{r^2}}{2}+\frac{\beta\ell^2}{2}-\frac{\displaystyle\int_{0}^{\infty}\dd{r}r^2\exp(-\beta\qty(\frac{\ell^2}{2r^2}+V(r)))}{\sqrt{\frac{2\pi}{\beta\ell^2}}\displaystyle\int_{0}^{\infty}\dd{r}r\ \erfc{\sqrt{\frac{\beta\ell^2}{2}\frac{1}{r}}}\exp(-\beta V(r))}.
\end{equation}
We reuse our notation $\widetilde{V}(r)=\ell^2/2r^2+V(r)$. We are now in the position to apply the saddle point approximation as $T\to0$. As earlier, we denote the strongest minimum of $\widetilde{V}(r)$ by $r_c$. We will use the asymptotic expansion of $\erfc{x}$ given by
\begin{equation}
    \erfc{x}=\frac{e^{-x^2}}{x\sqrt{\pi}}\sum_{n=0}^{\infty}(-1)^n\frac{(2n-1)!!}{(2x^2)^{n}}=\frac{e^{-x^2}}{x\sqrt{\pi}}\qty(1-\frac{1}{2x^2}+\cdots).
\end{equation}
The first two terms will suffice our needs. Firstly, we look at $\ev{r^2}$ as $\beta\to\infty$.
\begin{equation}
    \ev{r^2}=\frac{\displaystyle\int_{0}^{\infty}\dd{r}r^3\erfc{\sqrt{\frac{\beta\ell^2}{2}}\frac{1}{r}}\exp(-\beta V(r))}{\displaystyle\int_{0}^{\infty}\dd{r}r\ \erfc{\sqrt{\frac{\beta\ell^2}{2}}\frac{1}{r}}\exp(-\beta V(r))}\sim \frac{\sqrt{\frac{2}{\pi\beta\ell^2}}\displaystyle\int_{0}^{\infty}\dd{r}r^4\exp(-\beta \widetilde{V}(r))}{\sqrt{\frac{2}{\pi\beta\ell^2}}\displaystyle\int_{0}^{\infty}\dd{r}r^2\exp(-\beta \widetilde{V}(r))}.
\end{equation}
Now doing a saddle point approximation around $r=r_c$, and proceeding as we did in Sec. \ref{sec:sm_saddlepoint}, we get,
\begin{equation}
    \ev{r^2}(\beta\to\infty)=r_c^2.
\end{equation}
Next, we consider 
\begin{align*}
    \frac{\displaystyle\int_{0}^{\infty}\dd{r}r^2\exp(-\beta\qty(\frac{\ell^2}{2r^2}+V(r)))}{\sqrt{\frac{2\pi}{\beta\ell^2}}\displaystyle\int_{0}^{\infty}\dd{r}r\ \erfc{\sqrt{\frac{\beta\ell^2}{2}\frac{1}{r}}}\exp(-\beta V(r))}&\sim \frac{\displaystyle\int_{0}^{\infty}\dd{r}r^2\exp(-\beta\widetilde V(r))}{\sqrt{\frac{2\pi}{\beta\ell^2}}\sqrt{\frac{2}{\beta\ell^2\pi}}\displaystyle\int_{0}^{\infty}\dd{r}r\qty(r-\frac{r^3}{\beta\ell^2})\exp(-\beta\widetilde V(r))}\\
    &=\frac{1}{\frac{2}{\beta\ell^2}\qty(1-\frac{r_c^2}{\beta\ell^2})}\\
    &=\frac{\beta\ell^2}{2}\qty(1+\frac{r_c^2}{\beta\ell^2})\\
    \imp \frac{\displaystyle\int_{0}^{\infty}\dd{r}r^2\exp(-\beta\qty(\frac{\ell^2}{2r^2}+V(r)))}{\sqrt{\frac{2\pi}{\beta\ell^2}}\displaystyle\int_{0}^{\infty}\dd{r}r\ \erfc{\sqrt{\frac{\beta\ell^2}{2}\frac{1}{r}}}\exp(-\beta V(r))}&\overset{\beta\to\infty}{=} \frac{\beta\ell^2}{2}+\frac{r_c^2}{2}
\end{align*}
Thus,
\begin{equation}
    \ev{z^2}(\beta\to\infty)=\frac{r_c^2}{2}+\frac{\beta\ell^2}{2}-\qty(\frac{\beta\ell^2}{2}+\frac{r_c^2}{2})=0.
\end{equation}
Thus, $\Delta z(T\to0)=0$ and we have a trivial lower bound in both $\Delta z$ and $\Delta p_z$ with $\Delta z\Delta p_z\ge0$.

\subsection{Non-conserved Angular Momentum}
As in 2D, we consider the auxiliary Hamiltonian
\begin{equation}
    \widetilde H=H-\mu p_\phi=\frac{p_r^2}{2}+\frac{p_\theta^2}{2r^2}+\frac{p_\phi^2}{2r^2\sin^2\theta}+V(r)-\mu p_\phi
\end{equation}
with the Langevin dynamics given by
\begin{equation}
\begin{aligned}
    \dot r=p_r;~ \dot p_r=-\pdv{V}{r}+&\frac{p_\theta^2}{r^3}+\frac{p_\phi^2}{r^3\sin^2\theta};~ \dot\theta=\frac{p_\theta}{r^2};~ \dot p_\theta=\frac{p_\phi^2}{r^2}\cot\theta\csc^2\theta;~ \dot\phi=\frac{p_\phi}{r^2\sin^2\theta}-\mu;\\
    &\dot p_\phi=\mu r^2\gamma\sin^2\theta-\gamma p_\phi+r\sin\theta\sqrt{\gamma T}\xi(t)
    \end{aligned}
\end{equation}
with $\ev{\xi(t)}=0$ and $\ev{\xi(t)\xi(t')}=2\delta(t-t')$. The steady state is $P(r,p_r,\theta,p_\theta,p_\phi)=\frac{\exp(-\beta(H-\mu p_\phi))}{\Z(\beta,\mu)}$ where the partition function, $\Z(\beta,\mu)$, must converge for $P$ to a valid PDF. The partition function is given as 
\begin{align*}
    \Z(\beta,\mu)&=\int_{0}^{2\pi}\dd{\phi}\int_{0}^{\infty}\dd{r}\int_{0}^{\pi}\dd{\theta}\int_{-\infty}^{\infty}\dd{p_r}\int_{-\infty}^{\infty}\dd{p_\theta}\int_{-\infty}^{\infty}\dd{p_\phi}\exp(-\beta\qty(\frac{p_r^2}{2}+\frac{p_\theta^2}{2r^2}+\frac{p_\phi^2}{2r^2\sin^2\theta}+V(r)))\\
    &=\frac{4\pi^2}{\beta}\int_{0}^{\infty}\dd{r}r\int_{0}^{\pi}\dd{\theta}\int_{-\infty}^{\infty}\dd{p_\phi}\exp(-\frac{\beta}{2r^2\sin^2\theta}(p_\phi-\mu r^2\sin^2\theta)^2)\exp(\frac{\beta\mu^2r^2\sin^2\theta}{2})\exp(-\beta V(r))\\
    &=\frac{4\pi^2}{\beta}\sqrt{\frac{2\pi}{\beta}}\int_{0}^{\infty}\dd{r}r^2\int_{0}^{\pi}\dd{\theta}\sin\theta\exp(\frac{\beta\mu^2r^2\sin^2\theta}{2})\exp(-\beta V(r))\\
    \imp \Z(\beta,\mu)&=\frac{8\pi^3}{\beta^2\abs{\mu}}\int_{0}^{\infty}\dd{r}r\erf\qty(r\sqrt{\frac{\beta\mu^2}{2}})\exp(-\beta\qty(V(r)-\frac{\mu^2r^2}{2})). \numberthis
\end{align*}
As in the conserved case,
\begin{equation}
    \ev{x}=\ev{y}=\ev{z}=\ev{p_x}=\ev{p_y}=\ev{p_z}=0;\quad \ev{x^2}=\ev{y^2}=\frac{\ev{r^2\sin^2\theta}}{2},~~ \ev{z^2}=\frac{\ev{r^2\cos^2\theta}}{2}.
\end{equation}
Following a similar algebra as in the conserved case, it can be shown that
\begin{equation}
    \ev{\frac{p_\theta^2}{r^2}\sin^2\theta}=T\ev{\sin^2\theta}\imp \ev{p_z^2}=T\ev{\cos^2\theta}+\ev{\frac{p_\theta^2}{r^2}\sin^2\theta}=T\qty(\ev{\cos^2\theta}+\ev{\sin^2\theta})=T.
\end{equation}
We still have
\begin{equation}
    \ev{p_x^2}=\ev{p_y^2}=\frac{T}{2}+\ev{\frac{p_\phi^2}{2r^2\sin^2\theta}}.
\end{equation}
Now, we proceed as in the $2$D non-conserved angular momentum case in Sec. \ref{sec:sm_momentrelation} to find expressions of $\ev{p_\phi}$ and $\ev{\frac{p_\phi^2}{r^2\sin^2\theta}}$. 
\begin{align*}
    \ev{p_\phi}&=\frac{\displaystyle\int_{0}^{\infty}\dd{r}r\int_{0}^{\pi}\dd{\theta}\int_{-\infty}^{\infty}\dd{p_\phi} p_\phi\exp(-\frac{\beta}{2r^2\sin^2\theta}(p_\phi-\mu r^2\sin^2\theta)^2)\exp(\frac{\beta\mu^2r^2\sin^2\theta}{2})\exp(-\beta V(r))}{\displaystyle\int_{0}^{\infty}\dd{r}r\int_{0}^{\pi}\dd{\theta}\int_{-\infty}^{\infty}\dd{p_\phi}\exp(-\frac{\beta}{2r^2\sin^2\theta}(p_\phi-\mu r^2\sin^2\theta)^2)\exp(\frac{\beta\mu^2r^2\sin^2\theta}{2})\exp(-\beta V(r))}\\
    &=\frac{\displaystyle\int_{0}^{\infty}\dd{r}r\int_{0}^{\pi}\dd{\theta}\int_{-\infty}^{\infty}\dd{p_\phi} (p_\phi+\mu r^2\sin^2\theta)\exp(-\frac{\beta p_\phi^2}{2r^2\sin^2\theta})\exp(\frac{\beta\mu^2r^2\sin^2\theta}{2})\exp(-\beta V(r))}{\displaystyle\int_{0}^{\infty}\dd{r}r\int_{0}^{\pi}\dd{\theta}\int_{-\infty}^{\infty}\dd{p_\phi}\exp(-\frac{\beta p_\phi^2}{2r^2\sin^2\theta})\exp(\frac{\beta\mu^2r^2\sin^2\theta}{2})\exp(-\beta V(r))}\\
    \imp \ev{p_\phi}&=\mu\ev{r^2\sin^2\theta}. \numberthis
\end{align*}
Thus, if $\ev{r^2}$ exists, $\ev{r^2\sin^2\theta}$ must exist and we can always choose a $\mu$ for $p_\phi$ to fluctuate about any desired mean $\ev{p_\phi}$.  Similarly,
\begin{align*}
    \ev{\frac{p_\phi^2}{r^2\sin^2\theta}}&=\frac{\displaystyle\int_{0}^{\infty}\dd{r}r\int_{0}^{\pi}\frac{\dd{\theta}}{r^2\sin^2\theta}\int_{-\infty}^{\infty}\dd{p_\phi} p_\phi^2\exp(-\frac{\beta}{2r^2\sin^2\theta}(p_\phi-\mu r^2\sin^2\theta)^2)\exp(\frac{\beta\mu^2r^2\sin^2\theta}{2})e^{-\beta V(r)}}{\displaystyle\int_{0}^{\infty}\dd{r}r\int_{0}^{\pi}\dd{\theta}\int_{-\infty}^{\infty}\dd{p_\phi}\exp(-\frac{\beta}{2r^2\sin^2\theta}(p_\phi-\mu r^2\sin^2\theta)^2)\exp(\frac{\beta\mu^2r^2\sin^2\theta}{2})e^{-\beta V(r)}}\\
    &=\frac{\displaystyle\int_{0}^{\infty}\dd{r}r\int_{0}^{\pi}\frac{\dd{\theta}}{r^2\sin^2\theta}\int_{-\infty}^{\infty}\dd{p_\phi} (p_\phi^2+\mu^2 r^4\sin^4\theta)\exp(-\frac{\beta p_\phi^2}{2r^2\sin^2\theta})\exp(\frac{\beta\mu^2r^2\sin^2\theta}{2})e^{-\beta V(r)}}{\displaystyle\int_{0}^{\infty}\dd{r}r\int_{0}^{\pi}\dd{\theta}\int_{-\infty}^{\infty}\dd{p_\phi}\exp(-\frac{\beta p_\phi^2}{2r^2\sin^2\theta})\exp(\frac{\beta\mu^2r^2\sin^2\theta}{2})e^{-\beta V(r)}}
\end{align*}
The $p_\phi^2$ integral can be easily done and we end up with
\begin{equation}
    \ev{\frac{p_\phi^2}{r^2\sin^2\theta}}=T+\mu^2\ev{r^2\sin^2\theta}.
\end{equation}
Therefore, we have
\begin{equation}
    \ev{p_x^2}=\ev{p_y^2}=T+\frac{\mu^2}{2}\ev{r^2\sin^2\theta}=T+\frac{\ev{p_\phi}^2}{2\ev{r^2\sin^2\theta}}.
\end{equation}
Finally,
\begin{equation}
    \Delta x\Delta p_x=\Delta y\Delta p_y=\frac{1}{2}\sqrt{2T\ev{r^2\sin^2\theta}+\ev{p_\phi}^2}\geq \frac{\abs{\ev{p_\phi}}}{2}.
\end{equation}
Thus, we have the following uncertainty relations
\begin{equation}
    \Delta x\Delta p_x=\Delta y\Delta p_y\geq\frac{\abs{\ev{p_\phi}}}{2};\quad \Delta z\Delta p_z\geq0
\end{equation}
with the equality holding in the limit $T\to0$.
